\documentclass[12pt]{article}

\usepackage[cp1251]{inputenc}
\usepackage[T2A]{fontenc}
\usepackage[english,russian]{babel}
\usepackage{amsmath,amsfonts,amssymb}
\usepackage{amsbsy}
\usepackage{bm}

\usepackage{graphicx}
\usepackage{cite}
\usepackage{epstopdf}

\newtheorem{thm}{Теорема}

\newtheorem{Def}{Определение}

\renewcommand{\vec}[1]{\boldsymbol{#1}} 

\begin{document}

\begin{center}
\Large\textsc{\textbf{Application of the Kovacic algorithm for the investigation of motion of a heavy rigid body with a fixed point in the Hess case.}}
\end{center}
{\vskip 0.5cm}
\begin{center}
{\Large\bf Boris S. Bardin${}^{*}$, Alexander S. Kuleshov${}^{**}$}
{\vskip 0.3cm}
{${}^{*}$ Moscow Aviation Institute (national research university),\\
        Moscow 125080, Russia \textit{E-mail: bardin@yandex.ru}\\
        ${}^{**}$ M.~V.~Lomonosov Moscow State University,\\
        Moscow 119234. \textit{E-mail: kuleshov@mech.math.msu.su}}
\end{center}
{\vskip 1cm}

\begin{abstract}
In 1890 German mathematician and physicist W.~Hess~\cite{Hess} found new special case of integrability of Euler -- Poisson equations of motion of a heavy rigid body with a fixed point. In 1892 P.~A.~Nekrasov~\cite{Nekrasov1, Nekrasov2} proved that the solution of the problem of motion of a heavy rigid body with a fixed point under Hess conditions reduces to integrating the second order linear differential equation. In this paper the corresponding linear differential equation is derived and its coefficients are presented in the rational form. Using the Kovacic algorithm~\cite{Kovacic}, we proved that the liouvillian solutions of the corresponding second order linear differential equation exists only in the case, when the moving rigid body is the Lagrange top, or in the case when the constant of the area integral is zero.
\end{abstract}


\subsection{Euler -- Poisson equations}

Let us consider the problem of motion of a heavy rigid body with the fixed point $O$. To describe motion of the body we introduce two orthogonal coordinate systems: the fixed system $Oxyz$ and the moving system $Ox_1x_2x_3$. The $Oz$ axis of the fixed system is directed along the upward vertical. The $Ox_1x_2x_3$ system is rigidly connected with the moving body and its axes are directed along the principal axes of inertia at $O$. We denote the unit vectors of $Ox_1x_2x_3$ system by ${\bf e}_1$, ${\bf e}_2$,
${\bf e}_3$. Let $\vec\gamma=\gamma_1{\bf e}_1+\gamma_2{\bf e}_2+\gamma_3{\bf e}_3$ be the unit vector of the $Oz$ axis, where $\gamma_1$, $\gamma_2$, $\gamma_3$ are projections of this vector onto the $Ox_1x_2x_3$ axes. We denote by ${\bf P}$ the gravity force (directed vertically downward and applied at the center of gravity of the body $G$), then we have ${\bf P}=-Mg{\vec\gamma}$, where $M$ is the mass of the body and $g$ is the gravity acceleration. We will define the position of the center of mass of the body by the radius -- vector
\begin{equation*}
{\bf r}=\overrightarrow{OG}=x_1{\bf e}_1+x_2{\bf e}_2+x_3{\bf e}_3,
\end{equation*}
where $x_1$, $x_2$, $x_3$ are projections of this vector onto $Ox_1x_2x_3$ axes.

We will apply the principle of angular momentum, taking with respect to the point $O$, to derive the equations of motion of the body. The corresponding vector equation with respect to the fixed coordinate system $Oxyz$ takes the following form:
\begin{equation}\label{1}
\frac{d{\bf K}}{dt}={\bf M},
\end{equation}
where ${\bf M}$ is the total moment of the external forces (the force ${\bf P}$ in the considered case) about the fixed point $O$, and ${\bf K}$ is the angular momentum of the body with respect to the fixed point $O$. This vector can be represented as follows (see.~\cite[p.~331]{SyngeGriffith}):
\begin{equation}\label{2}
{\bf K}=\mathbb{J}_O\vec\omega,
\end{equation}
where $\vec\omega$ is the angular velocity vector of the body and $\mathbb{J}_O$ is the tensor of inertia of the body at $O$.

It is well known that the rate of change of the vector $\vec{K}$ -- the vector $d{\bf K}/dt$, calculated in the fixed coordinate system $Oxyz$, equals the rate of growth of this vector $\delta{\bf K}/{\delta t}$ (calculated in the moving coordinate system $Ox_1x_2x_3$) plus the rate of transport, i.e. (see~\cite[p.~347]{SyngeGriffith}):
\begin{equation}\label{3}
\frac{d{\bf K}}{dt}=\frac{\delta{\bf K}}{\delta t}+\left[\vec\omega\times{\bf K}\right].
\end{equation}

Taking into account~\eqref{3}, we can rewrite equation~\eqref{1} as follows:
\begin{equation}\label{4}
\frac{\delta{\bf K}}{\delta t}+\left[\vec\omega\times{\bf K}\right]=\left[{\bf r}\times{\bf P}\right].
\end{equation}

In the moving coordinate system $Ox_1x_2x_3$ we will write the vector $\vec\omega$, the vector ${\bf K}$ and the tensor $\mathbb{J}_O$ as follows:
\begin{equation*}
\vec\omega=\omega_1{\bf e}_1+\omega_2{\bf e}_2+\omega_3{\bf e}_3,
\end{equation*}
\begin{equation*}
\mathbb{J}_O=\left(
\begin{array}{lll}
A_1&0&0\\ \\
0&A_2&0\\ \\
0&0&A_3
\end{array}
\right),
\end{equation*}
\begin{equation*}
{\bf K}=\mathbb{J}_O\vec\omega=A_1\omega_1{\bf e}_1+A_2\omega_2{\bf e}_2+A_3\omega_3{\bf e}_3,
\end{equation*}
where $A_1$, $A_2$, $A_3$ -- are principal moments of inertia of the body at $O$.

If we take the projections of vector equation~\eqref{4} onto the $Ox_1x_2x_3$ axes we obtain the three scalar equations:
\begin{equation}\label{5}
\begin{array}{l}
A_1{\dot\omega}_1+\left(A_3-A_2\right)\omega_2\omega_3=Mg\left(x_3\gamma_2-x_2\gamma_3\right), \\ \\
A_2{\dot\omega}_2+\left(A_1-A_3\right)\omega_1\omega_3=Mg\left(x_1\gamma_3-x_3\gamma_1\right), \\ \\
A_3{\dot\omega}_3+\left(A_2-A_1\right)\omega_1\omega_2=Mg\left(x_2\gamma_1-x_1\gamma_2\right).
\end{array}
\end{equation}

These equations are called the Euler's equations of motion of a rigid body with a fixed point. These equations actually involve six parameters $A_1$, $A_2$, $A_3$, $x_1$, $x_2$, $x_3$, characterizing the mass distribution of the body relative to the principal axes of inertia at $O$, and six unknowns $\omega_1$, $\omega_2$, $\omega_3$, $\gamma_1$, $\gamma_2$, $\gamma_3$ to be solved as functions of time $t$. To be able to solve for them we need three more independent equations relating the $\omega_1$, $\omega_2$, $\omega_3$, $\gamma_1$, $\gamma_2$, $\gamma_3$. We have, since $\vec\gamma$ is a space -- fixed vector (i.e. it is fixed in the $Oxyz$ coordinate system), then
\begin{equation*}
\frac{\delta{\vec\gamma}}{\delta t}+\left[\vec\omega\times{\vec\gamma}\right]=0.
\end{equation*}

This vector equation gives the three scalar equations onto the $Ox_1x_2x_3$ axes
\begin{equation}\label{6}
\dot{\gamma}_1=\omega_3\gamma_2-\omega_2\gamma_3,\quad \dot{\gamma}_2=\omega_1\gamma_3-\omega_3\gamma_1,\quad
\dot{\gamma}_3=\omega_2\gamma_1-\omega_1\gamma_2.
\end{equation}

These equations are called the Poisson equations. Finally, we obtain the close system of six nonlinear differential equations~\eqref{5},~\eqref{6} with respect to the six unknown functions of time $t$: $\omega_1$, $\omega_2$, $\omega_3$, $\gamma_1$, $\gamma_2$, $\gamma_3$. These equations are called the Euler -- Poisson equations.

It is well known that to solve the Euler -- Poisson equations we need to find four independent autonomous first integrals of the system~\eqref{5},~\eqref{6}~\cite{Arhangelskii, Kozlov}. For any values of parameters $A_1$, $A_2$, $A_3$, $x_1$, $x_2$, $x_3$ of the body and for any initial conditions we have three first integrals of the Euler -- Poisson equations.
\begin{enumerate}
\item The energy integral $T+V=h$, where $T$ is the kinetic energy of the body, defined by the formula
\begin{equation*}
T=\frac{1}{2}\left(A_1\omega_1^2+A_2\omega_2^2+A_3\omega_3^2\right),
\end{equation*}
and $V$ is the potential energy of the body which has the form:
\begin{equation*}
V=Mg\left(x_1\gamma_1+x_2\gamma_2+x_3\gamma_3\right).
\end{equation*}

Thus with respect to the axes $Ox_1x_2x_3$ the energy integral takes the form:
\begin{equation*}
\frac{1}{2}\left(A_1\omega_1^2+A_2\omega_2^2+A_3\omega_3^2\right)+
Mg\left(x_1\gamma_1+x_2\gamma_2+x_3\gamma_3\right)=h.
\end{equation*}

\item The area integral
\begin{equation*}
\left({\bf K}\cdot\vec\gamma\right)=C,
\end{equation*}
which can be written with respect to the axes $Ox_1x_2x_3$ as follows:
\begin{equation*}
A_1\omega_1\gamma_1+A_2\omega_2\gamma_2+A_3\omega_3\gamma_3=C.
\end{equation*}

\item The geometrical integral
\begin{equation*}
\gamma_1^2+\gamma_2^2+\gamma_3^2=1.
\end{equation*}
\end{enumerate}

Thus for integrability of the Euler -- Poisson equations we need to find only one additional autonomous first integral. Under appropriate conditions on the parameters $A_1$, $A_2$, $A_3$, $x_1$, $x_2$, $x_3$ this integral can exist only in the following three cases~\cite{Arhangelskii, Kozlov}.
\begin{enumerate}
\item The Euler case $\left(x_1=x_2=x_3=0\right)$. The additional integral has the form:
\begin{equation*}
A_1^2\omega_1^2+A_2^2\omega_2^2+A_3^2\omega_3^2=k^2.
\end{equation*}

\item The Lagrange case $\left(A_1=A_2, x_1=x_2=0\right)$. In this case the additional integral has the form:
\begin{equation*}
\omega_3=\omega={\rm const}.
\end{equation*}

\item The Kovalevskaya case $\left(A_1=A_2=2A_3, x_2=x_3=0\right)$. If we denote
\begin{equation*}
\frac{Mgx_1}{A_3}=n,
\end{equation*}
then the additional first integral can be written as follows:
\begin{equation*}
\left(\omega_1^2-\omega_2^2-n\gamma_1\right)^2+\left(2\omega_1\omega_2-n\gamma_2\right)^2=j.
\end{equation*}
\end{enumerate}

There are no other general cases of integrability of Euler -- Poisson equations. However, there are several cases, when for the special initial conditions we can find the additional special integral. One of these cases is the Hess case which has been firstly described in 1890 by W.~Hess~\cite{Hess}. Let us consider this case in more details.

\subsection{The Hess case}

We suppose that the parameters of the body $A_1$, $A_2$, $A_3$, $x_1$, $x_2$, $x_3$ satisfy the conditions
\begin{equation}\label{7}
x_3=0,\qquad A_2\left(A_3-A_1\right)x_2^2=A_1\left(A_2-A_3\right)x_1^2,\qquad A_2\geq A_3\geq A_1.
\end{equation}

Let us prove that under conditions~\eqref{7} the Euler -- Poisson equations~\eqref{5},~\eqref{6} have an additional special integral (the Hess integral) of the form:
\begin{equation}\label{8}
A_1\omega_1 x_1+A_2\omega_2 x_2=0.
\end{equation}

To prove the existence of the integral~\eqref{8} we write two first equations of the system~\eqref{5} taking into account conditions~\eqref{7}:
\begin{equation}\label{9}
\begin{array}{l}
A_1{\dot\omega}_1+\left(A_3-A_2\right)\omega_2\omega_3=-Mgx_2\gamma_3, \\ \\
A_2{\dot\omega}_2+\left(A_1-A_3\right)\omega_1\omega_3=Mgx_1\gamma_3.
\end{array}
\end{equation}

Multiplying the first of equations~\eqref{9} by $x_1$ and the second by $x_2$ and taking their sum we obtain:
\begin{equation}\label{10}
\frac{d}{dt}\left(A_1\omega_1 x_1+A_2\omega_2 x_2\right)=\left(A_3-A_1\right)\omega_1\omega_3 x_2+\left(A_2-A_3\right)\omega_2\omega_3 x_1.
\end{equation}

The right hand part of equation~\eqref{10} can be transformed using~\eqref{7} to the form:
\begin{equation}\label{11}
\left(A_3-A_1\right)\omega_1\omega_3 x_2+\left(A_2-A_3\right)\omega_2\omega_3
x_1=\frac{\omega_3\left(A_2-A_3\right)x_1}{A_2x_2}\left(A_1\omega_1 x_1+A_2\omega_2 x_2\right).
\end{equation}

Taking into account equation~\eqref{11} we can conclude from the equation~\eqref{10}, that if at initial instant the equation~\eqref{8} is valid then it holds during the whole time of motion of the rigid body. Therefore under conditions~\eqref{7} the Euler -- Poisson equations~\eqref{5},~\eqref{6} possess the special integral~\eqref{8}.

For the first time this case of special integrability of Euler -- Poisson equations was discovered by W.~Hess~\cite{Hess}. Hess found this case of integrability when he searched for singular solutions of his own form of the Euler -- Poisson equation~\cite{Hess}. In 1892 the Hess case was rediscovered by G.~G.~Appelroth~\cite{Appelroth} when he analysed the branching of the general solution of Euler -- Poisson equations on the complex plane of time using the ideas of S.~V.~Kovalevskaya~\cite{Kowalevski1,Kowalevski2}. The detailed analytical investigation of the Hess case has been made by P.~A.~Nekrasov~\cite{Nekrasov1, Nekrasov2}. In his papers~\cite{Nekrasov1, Nekrasov2} Nekrasov presented both the Hess conditions and the Hess integral and reduced the solution of the problem to the integration of a second order linear differential equation. Nekrasov proved that in the Hess case the solution branches out on the complex plane of time. He investigated the analytical properties of the obtained second order linear differential equation and pointed out the basic properties of trajectories on the Poisson sphere. He proved also that on the zero level of the area integral the problem is integrable in elliptic functions. The Hess integral as well as the reduction to the second order linear differential equation was independently rediscovered in 1895 by Roger Liouville~\cite{Liouville}. The geometrical analysis and the modelling of the Hess top (in particular, on the zero level of the area integral) were given by N~E.~Zhukovsky~\cite{Zhukowski}. In the paper~\cite{Mlodzeevski} B.~K.~Mlodzeevskii and P.~A.~Nekrasov proved that under special restrictions solutions of the problem can have asymptotic behaviour.

The complete analysis of the phase trajectories in a Hess case was made by A.~M.~Kovalev~\cite{Kovalev1}. In the paper~\cite{Kovalev3} which substantially uses the results obtained in~\cite{Kovalev1}, the authors presented detailed classification of possible hodographs of the angular momentum vector in dependence of constants of the energy integral $h$ and the area integral $C$. In the paper~\cite{Kovalev2} the kinematic description of motion of a rigid body in the Hess case at zero level of the area integral was given by the hodographs method~\cite{Kharlamov1, Kharlamov2}. In the paper~\cite{Gashenenko} at zero level of the area integral the temporal and spatial evolution of the angular velocity vector and the angular moment vector was investigated. The motion of the body with a fixed point in the Hess case was represented by the rolling motion of the ellipsoid of inertia on a fixed plane.

Since the solution of the problem of motion of a heavy rigid body with a fixed point in the Hess case is reduced to solving the second order linear differential equation it is possible to set up the problem of existence of liouvillian solutions of the corresponding linear differential equation. For this purpose it is possible to apply the so-called Kovacic algorithm~\cite{Kovacic}, which allows to find liouvillian solutions of a second order linear differential equation in explicit form. If a linear differential equation has no liouvillian solutions, the Kovacic algorithm also allows one to ascertain this fact. The necessary condition for the application of the Kovacic algorithm to a second order linear differential equation is that the coefficients of this equation should be rational functions of independent variable.

Below we derive the second order linear differential equation in the Hess case and reduce its coefficients to the form of rational functions. Further we study the problem of existence of liouvillian solutions for the obtained second order linear differential equation using the Kovacic algorithm.

As the first step to derive the corresponding second order linear differential equation in the Hess case let us write the Euler -- Poisson equations in the special coordinate system, proposed by P.~V.~Kharlamov~\cite{Kharlamov1, Kharlamov2}.

\subsection{Euler -- Poisson equations in the special coordinate system}

Instead of the principal axes of inertia at $O$ (the $Ox_1x_2x_3$ coordinate system) let us choose the arbitrary trihedral coordinate system $O\xi_1\xi_2\xi_3$ which is rigidly connected with the moving body. We denote the unit vectors of the $O\xi_1\xi_2\xi_3$ system by ${\bf e}_I$, ${\bf e}_{II}$, ${\bf e}_{III}$. In this coordinate system we can write the angular momentum ${\bf K}$ as follows:
\begin{equation}\label{12}
{\bf K}=K_1{\bf e}_I+K_2{\bf e}_{II}+K_3{\bf e}_{III}.
\end{equation}

On the other hand, the angular momentum is defined by~\eqref{2}
\begin{equation*}
{\bf K}=\mathbb{J}_O\vec\omega,
\end{equation*}
where
\begin{equation}\label{13}
\vec\omega=\Omega_1{\bf e}_I+\Omega_2{\bf e}_{II}+\Omega_3{\bf e}_{III},
\end{equation}
\begin{equation}\label{14}
\mathbb{J}_O=\left(
\begin{array}{ccc}
L_{11}&-L_{12}&-L_{13}\\ \\
-L_{21}&L_{22}&-L_{23}\\ \\
-L_{31}&-L_{32}&L_{33}
\end{array}
\right).
\end{equation}

Here the components $L_{ij}$ of inertia tensor $\mathbb{J}_O$ satisfy the condition $L_{ij}=L_{ji}$. In this tensor components $L_{11}$, $L_{22}$, $L_{33}$ are the moments of inertia of the body with respect to $O\xi_1$, $O\xi_2$, $O\xi_3$ respectively and $L_{ij}$, $\left(i\ne j\right)$ are products of inertia. Using~\eqref{12},~\eqref{13},~\eqref{14} we can conclude that the angular momentum ${\bf K}$ has the following components onto $O\xi_1\xi_2\xi_3$ axes:
\begin{equation}\label{15}
\begin{array}{l}
K_1=L_{11}\Omega_1-L_{12}\Omega_2-L_{13}\Omega_3, \\ \\
K_2=-L_{21}\Omega_1+L_{22}\Omega_2-L_{23}\Omega_3, \\ \\
K_3=-L_{31}\Omega_1-L_{32}\Omega_2+L_{33}\Omega_3.
\end{array}
\end{equation}

We can write the kinetic energy of the moving body using the angular momentum ${\bf K}$ by the formula:
\begin{equation*}
T=\frac{1}{2}\left({\bf K}\cdot{\vec\omega}\right).
\end{equation*}

Taking into account~\eqref{13},~\eqref{15} the expression for the kinetic energy can be written in the scalar form:
\begin{equation}\label{16}
T=\frac{1}{2}\left(L_{11}\Omega_1^2+L_{22}\Omega_2^2+L_{33}\Omega_3^2\right)-L_{23}\Omega_2\Omega_3-L_{13}\Omega_1\Omega_3-
L_{12}\Omega_1\Omega_2.
\end{equation}

The inertia tensor~\eqref{14} is a non-singular matrix. Therefore equations~\eqref{15} can be represented as the linear nonhomogeneous equations with respect to $\Omega_1$, $\Omega_2$, $\Omega_3$. To solve equations~\eqref{15} with respect to $\Omega_1$, $\Omega_2$, $\Omega_3$, we find:
\begin{equation}\label{17}
\begin{array}{l}
\Omega_1=l_{11}K_1+l_{12}K_2+l_{13}K_3, \\ \\
\Omega_2=l_{21}K_1+l_{22}K_2+l_{23}K_3, \\ \\
\Omega_3=l_{31}K_1+l_{32}K_2+l_{33}K_3.
\end{array}
\end{equation}

Here we denote:
\begin{equation*}
l_{11}=\frac{L_{22}L_{33}-L_{23}^2}{\Delta},\qquad l_{22}=\frac{L_{11}L_{33}-L_{13}^2}{\Delta},\qquad
l_{33}=\frac{L_{11}L_{22}-L_{12}^2}{\Delta};
\end{equation*}
\begin{equation*}
l_{23}=l_{32}=\frac{L_{11}L_{23}+L_{12}L_{13}}{\Delta},\qquad l_{13}=l_{31}=\frac{L_{12}L_{23}+L_{13}L_{22}}{\Delta},\qquad
l_{12}=l_{21}=\frac{L_{12}L_{33}+L_{13}L_{23}}{\Delta};
\end{equation*}
\begin{equation*}
\Delta=\det\left(L_{ij}\right)=L_{11}L_{22}L_{33}-L_{11}L_{23}^2-L_{33}L_{12}^2-L_{22}L_{13}^2-2L_{12}L_{13}L_{23}.
\end{equation*}

Substituting~\eqref{17} in the formula~\eqref{16} for the kinetic energy $T$, we present it as the function of $K_1$, $K_2$, $K_3$:
\begin{equation}\label{18}
T=\frac{1}{2}\left(l_{11}K_1^2+l_{22}K_2^2+l_{33}K_3^2\right)+l_{23}K_2K_3+l_{13}K_1K_3+l_{12}K_1K_2.
\end{equation}

Now we simplify the obtained expression~\eqref{18} for the kinetic energy. For this purpose we rotate the coordinate system $O\xi_1\xi_2\xi_3$ about the $O\xi_1$ axis counter clockwise through an angle $\alpha$. We denote the obtained coordinate system by $O\xi_1^*\xi_2^*\xi_3^*$. In this case the new components of the vector ${\bf K}$, which we denote by $K_1^*$, $K_2^*$, $K_3^*$, can be expressed through the components $K_1$, $K_2$, $K_3$ by the following formulae:
\begin{equation*}
K_1=K_1^*, \qquad K_2=K_2^*\cos\alpha-K_3^*\sin\alpha,\qquad K_3=K_2^*\sin\alpha+K_3^*\cos\alpha.
\end{equation*}

The expression for the kinetic energy $T$ as a function of components $K_1^*$, $K_2^*$, $K_3^*$ of the vector ${\bf K}$ takes the form:
\begin{equation}\label{19}
T=\frac{1}{2}\left(l_{11}^*\left(K_1^*\right)^2+l_{22}^*\left(K_2^*\right)^2+l_{33}^*\left(K_3^*\right)^2\right)+l_{23}^*K_2^*K_3^*+
l_{13}^*K_1^*K_3^*+l_{12}^*K_1^*K_2^*.
\end{equation}

In this formula the coefficients $l_{ij}^*$ are expressed through the coefficients $l_{ij}$ and the angle $\alpha$ by the formulae:
\begin{equation*}
l_{11}^*=l_{11}, \quad l_{22}^*=l_{22}\cos^2\alpha+l_{33}\sin^2\alpha+l_{23}\sin 2\alpha,
\end{equation*}
\begin{equation*}
l_{33}^*=l_{22}\sin^2\alpha+l_{33}\cos^2\alpha-l_{23}\sin 2\alpha,
\end{equation*}
\begin{equation*}
l_{12}^*=l_{13}\sin\alpha+l_{12}\cos\alpha,\quad l_{13}^*=l_{13}\cos\alpha-l_{12}\sin\alpha,
\end{equation*}
\begin{equation*}
l_{23}^*=l_{23}\cos 2\alpha-\frac{1}{2}\left(l_{22}-l_{33}\right)\sin 2\alpha.
\end{equation*}

We assume now that the angle $\alpha$ is defined by the formula:
\begin{equation*}
\alpha=\frac{1}{2}\arctg\frac{2l_{23}}{l_{22}-l_{33}}.
\end{equation*}

In this case the coefficient $l_{23}^*$ in~\eqref{19} becomes zero and the expression for the kinetic energy takes the form:
\begin{equation}\label{20}
T=\frac{1}{2}\left(l_{11}^*\left(K_1^*\right)^2+l_{22}^*\left(K_2^*\right)^2+l_{33}^*\left(K_3^*\right)^2\right)+
\left(l_{13}^*K_3^*+l_{12}^*K_2^*\right)K_1^*.
\end{equation}

Assuming that the center of mass $G$ of the body does not coincide with the fixed point $O$ we choose the coordinate system $O\eta_1\eta_2\eta_3$ as follows. Let the $O\eta_1$ axis passes through the center of mass $G$ of the body (i.e. it is directed along the vector ${\bf r}$) and let the two other axes $O\eta_2$ and $O\eta_3$ be directed in such a way that the expression for the kinetic energy $T$ has a form~\eqref{20}, i.e. it does not contain the product $K_2^*K_3^*$. This coordinate system was firstly introduced by P.~V.~Kharlamov~\cite{Kharlamov1, Kharlamov2}. He called the $O\eta_1\eta_2\eta_3$ system as the special coordinate system. Obviously, this coordinate system is uniquely determined, and it will be rigidly connected with a rigid body. Using the special coordinate system we will write $a$, $a_1$, $a_2$, $b_1$, $b_2$ instead of $l_{11}^*$, $l_{22}^*$, $l_{33}^*$, $l_{12}^*$, $l_{13}^*$ and we will write $L_1$, $L_2$, $L_3$, $\omega_I$, $\omega_{II}$, $\omega_{III}$ instead of $K_1^*$, $K_2^*$, $K_3^*$, $\Omega_1$, $\Omega_2$, $\Omega_3$ respectively. Finally we have the following expression for the kinetic energy $T$ and components of the angular velocity $\vec\omega$ of the body onto $O\eta_1\eta_2\eta_3$ axes:
\begin{equation}\label{21}
T=\frac{1}{2}\left(aL_1^2+a_1L_2^2+a_2L_3^2\right)+\left(b_1L_2+b_2L_3\right)L_1.
\end{equation}
\begin{equation}\label{22}
\omega_I=aL_1+b_1L_2+b_2L_3,\qquad
\omega_{II}=a_1L_2+b_1L_1,\qquad
\omega_{III}=a_2L_3+b_2L_1.
\end{equation}

Using~\eqref{5},~\eqref{6},~\eqref{21},~\eqref{22} we can write the Euler -- Poisson equations in the special P.~V.~Kharlamov coordinate system as follows (see~\cite{Kharlamov1, Kharlamov2}):
\begin{equation}\label{23}
\begin{array}{l}
\dot{L}_1=\left(a_2-a_1\right)L_2L_3+\left(b_2L_2-b_1L_3\right)L_1,\\ \\
\dot{L}_2=\left(a-a_2\right)L_1L_3+\left(b_1L_2+b_2L_3\right)L_3-b_2L_1^2+\Gamma\nu_3,\\ \\
\dot{L}_3=-\left(a-a_1\right)L_1L_2-\left(b_1L_2+b_2L_3\right)L_2+b_1L_1^2-\Gamma\nu_2,\\ \\
\dot{\nu}_1=\left(a_2L_3+b_2L_1\right)\nu_2-\left(a_1L_2+b_1L_1\right)\nu_3,\\ \\
\dot{\nu}_2=\left(aL_1+b_1L_2+b_2L_3\right)\nu_3-\left(a_2L_3+b_2L_1\right)\nu_1,\\ \\
\dot{\nu}_3=-\left(aL_1+b_1L_2+b_2L_3\right)\nu_2+\left(a_1L_2+b_1L_1\right)\nu_1.
\end{array}
\end{equation}

Here $\nu_1$, $\nu_2$, $\nu_3$ are projections of the vector $\vec\gamma$ onto the $O\eta_1\eta_2\eta_3$ axes and $\Gamma=Mg\rho$, where $\rho=\sqrt{x_1^2+x_2^2+x_3^2}$.

We will use the special coordinate system $O\eta_1\eta_2\eta_3$ for the description of motion of a heavy rigid body with a fixed point in the Hess case.

\subsection{Equations of motion of a heavy rigid body with a fixed point in the Hess case written in the special coordinate system}

In the problem of motion of a heavy rigid body with a fixed point in the Hess case the transformation from the principal axes of inertia $Ox_1x_2x_3$ (the unit vectors ${\bf e}_1$, ${\bf e}_2$, ${\bf e}_3$) to the special P.~V.~Kharlamov coordinate system $O\eta_1\eta_2\eta_3$ (the unit vectors ${\bf e}_I$, ${\bf e}_{II}$, ${\bf e}_{III}$) is defined by the formulae:
\begin{equation*}
{\bf e}_I={\bf e}_1\cos\alpha+{\bf e}_2\sin\alpha, \qquad {\bf e}_{II}=-{\bf e}_1\sin\alpha+{\bf e}_2\cos\alpha,\qquad
{\bf e}_{III}={\bf e}_3,
\end{equation*}
where $\cos\alpha$ and $\sin\alpha$ equal
\begin{equation}\label{24}
\cos\alpha=\frac{x_1}{\sqrt{x_1^2+x_2^2}}, \qquad \sin\alpha=\frac{x_2}{\sqrt{x_1^2+x_2^2}}.
\end{equation}

Let us prove now that the unit vectors ${\bf e}_I$, ${\bf e}_{II}$, ${\bf e}_{III}$ are indeed the basis vectors of the special P.~V.~Kharlamov coordinate system. We note< first of all, that $x_3=0$ according to~\eqref{7}, and therefore the unit vector 
${\bf e}_I$ of the axis $O\eta_1$ is indeed collinear to the vector $\overrightarrow{OG}={\bf r}$. Now let us make sure that the kinetic energy of the body in this case is represented in the form~\eqref{21}. For this purpose we write the angular momentum as follows:
\begin{equation}\label{25}
{\bf K}=L_1{\bf e}_I+L_2{\bf e}_{II}+L_3{\bf e}_{III}=\left(L_1\cos\alpha-L_2\sin\alpha\right){\bf e}_1+
\left(L_1\sin\alpha+L_2\cos\alpha\right){\bf e}_2+L_3{\bf e}_3.
\end{equation}

In the principal axes of inertia the angular momentum has the form
\begin{equation}\label{26}
{\bf K}=K_1{\bf e}_1+K_2{\bf e}_2+K_3{\bf e}_3=A_1\omega_1{\bf e}_1+A_2\omega_2{\bf e}_2+A_3\omega_3{\bf e}_3,
\end{equation}
and the kinetic energy of the body can be written as follows:
\begin{equation}\label{27}
T=\frac{1}{2}\left(\frac{K_1^2}{A_1}+\frac{K_2^2}{A_2}+\frac{K_3^2}{A_3}\right).
\end{equation}

It follows from~\eqref{25},~\eqref{26} that the components $L_1$, $L_2$, $L_3$ of the angular momentum ${\bf K}$ with respect to the $O\eta_1\eta_2\eta_3$ coordinate system are connected with the components $K_1$, $K_2$, $K_3$ of this vector with respect to the principal axes of inertia by the formulae:
\begin{equation}\label{28}
K_1=L_1\cos\alpha-L_2\sin\alpha,\qquad K_2=L_1\sin\alpha+L_2\cos\alpha,\qquad K_3=L_3.
\end{equation}

If we substitute expressions~\eqref{28} to the formula~\eqref{27} for the kinetic energy of the body and take into account the explicit expressions~\eqref{24} for $\sin\alpha$ and $\cos\alpha$, we obtain the following expression for the kinetic energy of the body as a function of $L_1$, $L_2$, $L_3$:
\begin{equation}\label{29}
T\!=\!\frac{1}{2}\left(\frac{x_1^2}{A_1}\!+\!\frac{x_2^2}{A_2}\right)\frac{L_1^2}{\left(x_1^2\!+\!x_2^2\right)}\!+\!\frac{1}{2}
\left(\frac{x_1^2}{A_2}\!+\!\frac{x_2^2}{A_1}\right)\frac{L_2^2}{\left(x_1^2\!+\!x_2^2\right)}\!+\!\frac{1}{2}\frac{L_3^2}{A_3}\!+\!
\left(\frac{1}{A_2}\!-\!\frac{1}{A_1}\right)\frac{x_1x_2}{\left(x_1^2\!+\!x_2^2\right)}L_1L_2.
\end{equation}

Note that expression~\eqref{29} for the kinetic energy of the rigid body does not contain the products $L_1L_3$ and $L_2L_3$. This means, that the kinetic energy of the body in the considered case has the form~\eqref{21}, i.e. the unit vectors ${\bf e}_I$, ${\bf e}_{II}$, ${\bf e}_{III}$ are indeed the basis vectors of the special coordinate system by P.~V.~Kharlamov~\cite{Kharlamov1, Kharlamov2}. The coefficients of the kinetic energy~\eqref{29} are such, that
\begin{equation*}
b_2=0,\quad l_{23}^*=0.
\end{equation*}

Moreover, according to~\eqref{7} we have:
\begin{equation*}
\frac{A_1x_1^2+A_2x_2^2}{A_1A_2\left(x_1^2+x_2^2\right)}=\frac{1}{A_3},
\end{equation*}
and therefore
\begin{equation*}
a_1=a_2.
\end{equation*}

Finally expression~\eqref{29} for the kinetic energy of the body with a fixed point in the Hess case can be written as follows:
\begin{equation}\label{30}
T=\frac{1}{2}aL_1^2+\frac{1}{2}c\left(L_2^2+L_3^2\right)+bL_1L_2,
\end{equation}
where we denote
\begin{equation*}
a=\frac{A_2x_1^2+A_1x_2^2}{A_1A_2\left(x_1^2+x_2^2\right)},\qquad
b=\frac{\left(A_1-A_2\right)x_1x_2}{A_1A_2\left(x_1^2+x_2^2\right)},\qquad c=\frac{1}{A_3}.
\end{equation*}

Taking into account expression~\eqref{30} for the kinetic energy we can write the Euler -- Poisson equations~\eqref{23} in the form:
\begin{equation}\label{31}
\begin{array}{l}
\dot{L}_1=-bL_1L_3,\\ \\
\dot{L}_2=\left(a-c\right)L_1L_2+bL_2L_3+\nu_3\Gamma,\\ \\
\dot{L}_3=-\left(a-c\right)L_1L_2+bL_1^2-bL_2^2-\nu_2\Gamma,\\ \\
\dot{\nu}_1=cL_3\nu_2-\left(cL_2+bL_1\right)\nu_3,\\ \\
\dot{\nu}_2=-cL_3\nu_1+\left(aL_1+bL_2\right)\nu_3,\\ \\
\dot{\nu}_3=\left(bL_1+cL_2\right)\nu_1-\left(aL_1+bL_2\right)\nu_2.
\end{array}
\end{equation}

To find the additional first integral, existing in the Hess case, we consider the first equation of the system~\eqref{31}
\begin{equation*}
\dot{L}_1=-bL_1L_3.
\end{equation*}

In this equation, the right -- hand side is equal to the variable $L_1$ itself, multiplied by the coefficient $-bL_3$ bounded in absolute value. This means that if at the initial instant of time the quantity $L_1=0 $, then we have
\begin{equation}\label{32}
L_1\equiv 0.
\end{equation}

The invariant manifold~\eqref{32} (or, in other notations~\eqref{8}) together with~\eqref{7} defines the Hess case. Under conditions~\eqref{7},~\eqref{32} equations~\eqref{31} are noticeably simplified and take the form
\begin{equation}\label{33}
\begin{array}{c}
\dot{L}_2=bL_2L_3+\nu_3\Gamma,\quad \dot{L}_3=-bL_2^2-\nu_2\Gamma,\\ \\
\dot{\nu}_1=cL_3\nu_2-cL_2\nu_3,\quad \dot{\nu}_2=bL_2\nu_3-cL_3\nu_1,\quad
\dot{\nu}_3=cL_2\nu_1-bL_2\nu_2.
\end{array}
\end{equation}

Equations~\eqref{33} possess the following first integrals:
\begin{equation}\label{34}
\displaystyle\frac{c}{2}\left(L_2^2+L_3^2\right)+\Gamma\nu_1=E;\quad L_2\nu_2+L_3\nu_3=k;\quad
\nu_1^2+\nu_2^2+\nu_3^2=1.
\end{equation}

Note that condition $b=0$ corresponds to the Lagrange integrable case in the problem of motion of a heavy rigid body with a fixed point.

\subsection{Dimensionless equations. Transformation to the second-order linear differential equation}

Now let us write the equations~\eqref{33} and the first integrals~\eqref{34} in dimensionless form. For this purpose we introduce the dimensionless components of angular momentum
\begin{equation*}
L_2=\sqrt{\frac{\Gamma}{c}}y,\qquad L_3=\sqrt{\frac{\Gamma}{c}}z,
\end{equation*}
and the dimensionless time $\tau$:
\begin{equation*}
t=\frac{\tau}{\sqrt{\Gamma c}}.
\end{equation*}

We introduce also the dimensionless parameter
\begin{equation*}
d_1=\frac{b}{c}.
\end{equation*}
and the dimensionless constants of the first integrals
\begin{equation*}
h=\frac{E}{\Gamma},\qquad k_1=k\sqrt{\frac{c}{\Gamma}}.
\end{equation*}

Now we can write equations~\eqref{33} in dimensionless form:
\begin{equation}\label{35}
\begin{array}{c}
\displaystyle\frac{dy}{d\tau}=d_1yz+\nu_3,\qquad \displaystyle\frac{dz}{d\tau}=-d_1y^2-\nu_2,\\ \\
\displaystyle\frac{d{\nu}_1}{d\tau}=z\nu_2-y\nu_3,\quad \displaystyle\frac{d{\nu}_2}{d\tau}=d_1y\nu_3-z\nu_1,\quad
\displaystyle\frac{d{\nu}_3}{d\tau}=y\nu_1-d_1y\nu_2.
\end{array}
\end{equation}

System~\eqref{35} possesses the following first integrals:
\begin{equation}\label{36}
\frac{y^2+z^2}{2}+\nu_1=h,\quad y\nu_2+z\nu_3=k_1,\quad \nu_1^2+\nu_2^2+\nu_3^2=1.
\end{equation}

From the system~\eqref{35} using~\eqref{36} we will obtain the second order linear differential equation. Before we obtain this equation, let us determine the range of parameters $d_1$, $h$, $k_1$. It is easy to see that the parameter $k_1$ ranges in the infinite interval $\left(-\infty,\, +\infty\right)$. Since the expression
\begin{equation*}
\frac{y^2+z^2}{2}
\end{equation*}
is nonnegative, then we have for the parameter $h$ the following inequality
\begin{equation*}
h-\nu_1\geq 0,\quad\mbox{or}\quad h\geq\nu_1.
\end{equation*}

The minimal value of the first component $\nu_1$ of the vector $\vec\nu$ equals $-1$. Therefore the parameter $h$ ranges in the interval
\begin{equation*}
h\in\left[-1,\;+\infty\right).
\end{equation*}

The parameter $d_1$ can be represented as follows:
\begin{equation*}
d_1=\frac{b}{c}=\frac{\left(A_1-A_2\right)x_1x_2}{\left(A_1x_1^2+A_2x_2^2\right)}.
\end{equation*}

First of all we note that $d_1<0$ according to~\eqref{7}. We transform the expression for $d_1$ to the form:
\begin{equation}\label{37}
d_1=\frac{\left(A_1-A_2\right)x_1x_2}{\left(A_1x_1^2+A_2x_2^2\right)}=\frac{\left(A_1-A_2\right)\frac{x_1}{x_2}}
{\left(A_1\frac{x_1^2}{x_2^2}+A_2\right)}.
\end{equation}

From the conditions~\eqref{7} of existence of the Hess integral, we find
\begin{equation*}
\frac{x_1^2}{x_2^2}=\frac{A_2\left(A_3-A_1\right)}{A_1\left(A_2-A_3\right)},\quad\mbox{то есть}\quad
\frac{x_1}{x_2}=\frac{\sqrt{A_2\left(A_3-A_1\right)}}{\sqrt{A_1\left(A_2-A_3\right)}}.
\end{equation*}

Using this equation we can transform expression~\eqref{37} for the parameter $d_1$ as follows:
\begin{equation*}
d_1=-\sqrt{\frac{\left(A_2-A_3\right)\left(A_3-A_1\right)}{A_1A_2}},\quad\mbox{or}\quad
d_1^2=\frac{\left(A_2-A_3\right)\left(A_3-A_1\right)}{A_1A_2}.
\end{equation*}

Since the moments of inertia satisfy the triangle inequality
\begin{equation*}
A_1+A_3>A_2,\quad\mbox{i.e.}\quad\frac{A_2-A_3}{A_1}<1,
\end{equation*}
\begin{equation*}
A_1+A_2>A_3,\quad\mbox{i.e.}\quad\frac{A_3-A_1}{A_2}<1,
\end{equation*}
then taking into account these inequalities we have
\begin{equation*}
d_1^2=\frac{\left(A_2-A_3\right)}{A_1}\cdot\frac{\left(A_3-A_1\right)}{A_2}<1.
\end{equation*}

Since $d_1<0$ then we finally obtain that the parameter $d_1$ ranges in the interval
\begin{equation*}
d_1\in\left(-1,\, 0\right].
\end{equation*}

We obtain now the second order linear differential equation from the system~\eqref{35} using~\eqref{36}. Multiplying the first equation of the system~\eqref{35} by $y$ and the second by $z$ and adding them, we get:
\begin{equation}\label{38}
\frac{d}{d\tau}\left(\frac{y^2+z^2}{2}\right)=y\nu_3-z\nu_2.
\end{equation}

Using the following identity
\begin{equation*}
\left(y^2+z^2\right)\left(\nu_2^2+\nu_3^2\right)=\left(y\nu_2+z\nu_3\right)^2+\left(y\nu_3-z\nu_2\right)^2,
\end{equation*}
we find from the first integrals~\eqref{36}:
\begin{equation*}
\nu_1=h-\frac{y^2+z^2}{2}.
\end{equation*}

Therefore we have
\begin{equation*}
\begin{array}{l}
\nu_2^2+\nu_3^2=1-\left(h-\displaystyle\frac{y^2+z^2}{2}\right)^2=1-\left(\displaystyle\frac{y^2+z^2}{2}-h\right)^2,\\ \\
y\nu_2+z\nu_3=k_1.
\end{array}
\end{equation*}

Finally we obtain
\begin{equation*}
\left(y\nu_3-z\nu_2\right)^2=\left(y^2+z^2\right)\left(1-\left(\frac{y^2+z^2}{2}-h\right)^2\right)-k_1^2.
\end{equation*}

We will take that
\begin{equation}\label{39}
y\nu_3-z\nu_2=-\sqrt{\left(y^2+z^2\right)\left(1-\left(\frac{y^2+z^2}{2}-h\right)^2\right)-k_1^2}
\end{equation}
(we can choose the arbitrary sign before the square root in~\eqref{39}). Taking into account~\eqref{39} we can rewrite~\eqref{38} in the form:
\begin{equation*}
\frac{d}{d\tau}\left(\frac{y^2+z^2}{2}\right)=-\sqrt{\left(y^2+z^2\right)\left[1-\left(\frac{y^2+z^2}{2}-h\right)^2\right]-k_1^2}.
\end{equation*}

Now we multiply the first equation of the system~\eqref{35} by $z$ and the second -- by $y$ and subtract the first equation from the second equation. As a result we obtain:
\begin{equation*}
y\frac{dz}{d\tau}-z\frac{dy}{d\tau}=-d_1y^3-y\nu_2-d_1yz^2-z\nu_3,
\end{equation*}
or, taking into account~\eqref{36}
\begin{equation*}
y\frac{dz}{d\tau}-z\frac{dy}{d\tau}=-d_1y\left(y^2+z^2\right)-k_1.
\end{equation*}

We pass now from the variables $y$ and $z$ to the polar coordinates $x$ and $\varphi$ by putting:
\begin{equation*}
y=x\cos\varphi,\quad z=x\sin\varphi.
\end{equation*}

Then for the variables $x$ and $\varphi$ we have the following system of two differential equations:
\begin{equation}\label{40}
\begin{array}{l}
x\displaystyle\frac{dx}{d\tau}=-\sqrt{x^2\left[1-\left(\displaystyle\frac{x^2}{2}-h\right)^2\right]-k_1^2}, \\ \\
x^2\displaystyle\frac{d\varphi}{d\tau}=-d_1x^3\cos\varphi-k_1.
\end{array}
\end{equation}

From this system we obtain the single first order differential equation for the function $\varphi=\varphi\left(x\right)$:
\begin{equation}\label{41}
\frac{d\varphi}{dx}=\frac{d_1 x^3\cos\varphi+k_1}{x\sqrt{x^2\left[1-\left(
\displaystyle\frac{x^2}{2}-h\right)^2\right]-k_1^2}}.
\end{equation}

Note that when we pass from the system~\eqref{40} to the equation~\eqref{41} we exclude the case $x={\rm const}$ that is, $y^2+z^2={\rm const}$ or $\nu_1={\rm const}$ from consideration. Meanwhile for a heavy rigid body with a fixed point in the Hess case there are steady motions for which $\nu_1=\nu_1^0={\rm const}$ (see, for example~\cite{Novikov}).

The substitution
\begin{equation*}
w=\tan\frac{\varphi}{2}
\end{equation*}
reduces~\eqref{41} to the Riccati equation:
\begin{equation*}
\frac{dw}{dx}+\frac{\left(d_1x^3-k_1\right)w^2-d_1x^3-k_1}{2x\sqrt{x^2\left[1-\left(
\displaystyle\frac{x^2}{2}-h\right)^2\right]-k_1^2}}=0.
\end{equation*}
or similarly
\begin{equation*}
\frac{dw}{dx}=-\frac{d_1 x^3-k_1}{2x\sqrt{x^2\left[1-\left(
\displaystyle\frac{x^2}{2}-h\right)^2\right]-k_1^2}}\, w^2+\frac{d_1 x^3+k_1}{2x\sqrt{x^2\left[1-\left(
\displaystyle\frac{x^2}{2}-h\right)^2\right]-k_1^2}}.
\end{equation*}

It is well known from the general theory of ordinary differential equations (see, for example~\cite{ZaitsevPolyanin}), that if the Riccati equation has the form:
\begin{equation*}
\frac{dw}{dx}=f_2\left(x\right)w^2+f_1\left(x\right)w+f_0\left(x\right),
\end{equation*}
then the substitution of the form
\begin{equation*}
u\left(x\right)=\exp\left(-\int f_2\left(x\right)w\left(x\right)dx\right)
\end{equation*}
reduces it to the second order linear differential equation
\begin{equation}\label{42}
f_2\frac{d^2u}{dx^2}-\left(\frac{df_2}{dx}+f_1f_2\right)\frac{du}{dx}+f_0f_2^2u=0
\end{equation}
or, if we divide this equation by $f_2$:
\begin{equation}\label{43}
\frac{d^2u}{dx^2}-\left(\frac{1}{f_2}\frac{df_2}{dx}+f_1\right)\frac{du}{dx}+f_0f_2u=0.
\end{equation}

In our case
\begin{equation*}
f_2=-\frac{d_1x^3-k_1}{2x\sqrt{x^2\left[1-\left(
\displaystyle\frac{x^2}{2}-h\right)^2\right]-k_1^2}},\quad f_1=0,\quad
f_0=\frac{d_1x^3+k_1}{2x\sqrt{x^2\left[1-\left(
\displaystyle\frac{x^2}{2}-h\right)^2\right]-k_1^2}}.
\end{equation*}

Note that the transition from the equation~\eqref{42} to the equation~\eqref{43} is possible only when $f_2\ne 0$. Taking into account the fact that $x\ne {\rm const}$, the condition $f_2=0$ is equivalent to the simultaneous fulfillment of the conditions
\begin{equation}\label{44}
d_1=0,\quad k_1=0.
\end{equation}

Under the conditions~\eqref{44}, the equation~\eqref{41} gives $\varphi=\varphi_0={\rm const}$. It can be shown (see Appendix) that, under the conditions~\eqref{44} a heavy rigid body with a fixed point in the Hess case will perform pendulum nutational oscillations. We will assume further that $f_2\ne 0$. Thus, the problem of motion of a heavy rigid body with a fixed point in the Hess case is reduced to solving the following second order linear differential equation with the rational coefficients:
\begin{equation}\label{45}
\frac{d^2u}{dx^2}+a\left(x\right)\frac{du}{dx}+b\left(x\right)u=0,
\end{equation}
\begin{equation*}
a\left(x\right)=\frac{d_1x^9-4k_1x^6-4d_1\left(h^2-1\right)x^5+12k_1hx^4-8k_1^2d_1x^3-8k_1\left(h^2-1\right)x^2-4k_1^3}{x\left(
x^6-4hx^4+4\left(h^2-1\right)x^2+4k_1^2\right)\left(d_1x^3-k_1\right)},
\end{equation*}
\begin{equation*}
b\left(x\right)=\frac{\left(d_1x^3+k_1\right)\left(d_1x^3-k_1\right)}{x^2\left(x^6-4hx^4+4\left(h^2-1\right)x^2+4k_1^2\right)}.
\end{equation*}

Now we can study the problem of existence of liouvillian solutions for the second order linear differential equation~\eqref{45}. To solve this problem we can use the Kovacic algorithm~\cite{Kovacic}. Below we give a brief description of this algorithm.

\subsection{Description of the Kovacic algorithm}

Let us consider the differential field $\mathbb{C}\left(x\right)$ of rational functions of one (in general case complex) variable $x$. We accept the standard notations $\mathbb{Z}$ and $\mathbb{Q}$ for the sets of integer and rational numbers respectively. Our goal is to find a solution of the differential equation
\begin{equation}\label{46}
\frac{d^2z}{dx^2}+a\left(x\right)\frac{dz}{dx}+b\left(x\right)z=0,
\end{equation}
where $a\left(x\right), b\left(x\right)\in \mathbb{C}\left(x\right)$. In the paper~\cite{Kovacic} an algorithm has been proposed that allows one to find explicitly the so-called liouvillian solutions of differential equation~\eqref{46}, i.e. solutions, that can be expressed in terms of liouvillian functions. The main advantage of the Kovacic algorithm is precisely that it allows one not only to establish the existence or nonexistence of a solution of differential equation~\eqref{46} expressed in terms of liouvillian functions, but also to present this solution in an explicit form when it exists. In turn, liouvillian functions are elements of a liouvillian field, which is defined in the following way.

\begin{Def}
Let $F$ be a differential field of functions of one (in general case complex) variable $x$ that contains $\mathbb{C}\left(x\right)$; namely $F$ is a field of characteristic zero with a differentiation operator $\left(\right)'$ with the following two properties: $\left(a+b\right)'=a'+b'$ and $\left(ab\right)'=a'b+ab'$ for any $a$ and $b$ in $F$. The field $F$ is liouvillian if there exists a sequence (tower) of differential fields
\begin{equation*}
\mathbb{C}\left(x\right)=F_0\subseteq F_1\subseteq\ldots\subseteq F_n=F,
\end{equation*}
obtained by adjoining one element such that for any $i~=~1,2,\ldots, n$ we have:
\begin{equation*}
F_i=F_{i-1}\left(\alpha\right),\;\mbox{with}\;\, \frac{\alpha'}{\alpha}\in F_{i-1}
\end{equation*}
(i.e. $F_i$ is generated by an exponential of an indefinite integral over $F_{i-1}$); or
\begin{equation*}
F_i=F_{i-1}\left(\alpha\right),\;\mbox{with}\;\, \alpha'\in F_{i-1}
\end{equation*}
(i.e. $F_i$ is generated by an indefinite integral over $F_{i-1}$); or $F_i$ is finite algebraic over $F_{i-1}$ (i.e. $F_i=F_{i-1}\left(\alpha\right)$ and $\alpha$ satisfies a polynomial equation of the form
\begin{equation*}
a_0+a_1\alpha+\cdots+a_n\alpha^n=0,
\end{equation*}
where $a_j\in F_{i-1}$, $j=0,1,2,\ldots, n$ and are not all zero).
\end{Def}

Thus, liouvillian functions are built up sequentially from rational functions by using algebraic operations and the operation of indefinite integration and by taking the exponential of a given expression. A solution of equation~\eqref{46} that is expressed in terms of liouvillian functions most closely correspond to the notion of a "close-form solution"$\,$ or a "solution in quadratures". 
To reduce differential equation~\eqref{46} to a simpler form, we use the following formula
\begin{equation}\label{47}
y\left(x\right)=z\left(x\right)\exp\left(\frac{1}{2}\int a(x)dx\right).
\end{equation}

Then equation~\eqref{46} takes the form:
\begin{equation}\label{48}
y''=R\left(x\right)y,\quad R\left(x\right)=\frac{1}{2}a'+\frac{1}{4}a^2-b,\quad R\left(x\right)\in\mathbb{C}\left(x\right).
\end{equation}

Hereinafter, it is assumed that that the second order linear differential equation with which the Kovacic algorithm deals is written in the form~\eqref{48}. The following theorem which has been proved by J.~Kovacic~\cite{Kovacic}, determines the structure of a solution of this differential equation.
\begin{thm}\label{Theorem1}
For the differential equation~\eqref{48} only the following four cases are true.
\begin{enumerate}
\item The differential equation~\eqref{48} has a solution of the form
\begin{equation*}
\eta=\exp\left(\int\omega(x)dx\right)\quad\mbox{e}\quad\omega\left(x\right)\in\mathbb{C}\left(x\right)
\end{equation*}
(liouvillian solution of type $1$).

\item The differential equation~\eqref{48} has a solution of the form
\begin{equation*}
\eta=\exp\left(\int\omega(x)dx\right),
\end{equation*}
where $\omega\left(x\right)$ is an algebraic function of degree $2$ over $\mathbb{C}\left(x\right)$ and case $1$ does not hold (liouvillian solution of type $2$).

\item All solutions of differential equation~\eqref{48} are algebraic over $\mathbb{C}\left(x\right)$ and cases $1$ and $2$ do not hold. In this situation a solution of the differential equation~\eqref{48} has the form
\begin{equation*}
\eta=\exp\left(\int\omega(x)dx\right)
\end{equation*}
where $\omega\left(x\right)$ is an algebraic function of degree $4$, $6$ or $12$ over $\mathbb{C}\left(x\right)$ (liouvillian solution of type $3$).

\item Differential equation~\eqref{48} has no liouvillian solutions.
\end{enumerate}
\end{thm}

In order for one of the first three cases listed in Theorem~\ref{Theorem1} to take place the function $R\left(x\right)$ in the right hand side of equation~\eqref{48} must satisfy certain conditions. These conditions are necessary but not sufficient. For example, if the conditions corresponding to Case 1 of Theorem~\ref{Theorem1} are violated, then we must turn to the verification of the conditions corresponding to Cases 2 and 3. If these conditions are fulfilled, then we must search for solutions of equation~\eqref{48} exactly in the form, indicated for the corresponding case. However, the existence of such a solution is not guaranteed. In order to explain the sense of the necessary conditions mentioned, we recall some facts from complex analysis.

Recall that any analytic function $f$ of a complex variable $z$ can be expanded in a Laurent series in a neighborhood of any point $a$ as follows:
\begin{equation*}
f\left(z\right)=a_0+a_1\left(z-a\right)+a_2\left(z-a\right)^2+\cdots+\frac{a_{-1}}{z-a}+\frac{a_{-2}}{\left(z-a\right)^2}+\cdots.
\end{equation*}

The part of this series 
\begin{equation*}
a_0+a_1\left(z-a\right)+a_2\left(z-a\right)^2+\cdots
\end{equation*}
containing nonnegative powers of $z-a$ is called the analytic part of the Laurent series. Whereas the other part, namely
\begin{equation*}
\frac{a_{-1}}{z-a}+\frac{a_{-2}}{\left(z-a\right)^2}+\cdots
\end{equation*}
is called the principal part of the expansion. By definition, a point $a$ is called a pole of $f\left(z\right)$ of order $n$ if the principal part of the Laurent expansion contains a finite number of terms and the last term has the form 
\begin{equation*}
\frac{a_{-n}}{\left(z-a\right)^n}.
\end{equation*}

If $f\left(z\right)$ is a rational function of $z$, then a point $a$ is a pole of $f\left(z\right)$ of order $n$ if it is a root of the denominator of $f\left(z\right)$ of multiplicity $n$. 

Let $z=\infty$ be a zero of a function $f\left(z\right)$ of order $n$ (i.e., $n$ is the order of the pole at $z=0$ of $f\left(z\right)$). Then we say that $n$ is the order of $f\left(z\right)$ at $z=\infty$. If $f\left(z\right)$ is a rational function, then its order at $z=\infty$ is the difference between the degrees of the denominator and the numerator.

The following theorem, which has been proved in~\cite{Kovacic}, specifies conditions, that are necessary for one of the first three cases listed in Theorem~\ref{Theorem1} can hold.

\begin{thm}\label{Theorem2}
For the differential equation~\eqref{48} the following conditions are necessary for one of the first cases listed in Theorem~\ref{Theorem1} to hold, i.e. for equation~\eqref{48} to have a liouvillian solution of the type specified in description of the corresponding case.
\begin{enumerate}
\item Each pole of the function $R\left(x\right)$ must have even order or else have order $1$. The order of $R\left(x\right)$ at $x=\infty$ must be even or else be greater than 2.
\item The function $R\left(x\right)$ must have at least one pole that either has odd order greater than $2$ or else has order $2$.
\item The order of a pole of $R\left(x\right)$ cannot exceed $2$ and the order of $R\left(x\right)$ at $x=\infty$ must be at least $2$. If the partial fraction expansion of $R\left(x\right)$ has the form
\begin{equation*}
R\left(x\right)=\sum\limits_i\frac{\alpha_i}{\left(x-c_i\right)^2}+\sum\limits_j\frac{\beta_j}{x-d_j},
\end{equation*}
then for each $i$
\begin{equation*}
\sqrt{1+4\alpha_i}\in\mathbb{Q},\quad \sum\limits_j\beta_j=0
\end{equation*}
and if
\begin{equation*}
\gamma=\sum\limits_i\alpha_i+\sum\limits_j\beta_j d_j,
\end{equation*}
then
\begin{equation*}
\sqrt{1+4\gamma}\in\mathbb{Q}.
\end{equation*}
\end{enumerate}
\end{thm}

To find a liouvillian solution of type 1 of the differential equation~\eqref{48}, the Kovacic algorithm is stated in the following way (see~\cite{Kovacic} for details). We assume that the necessary conditions for the existence of a solution in case 1 are satisfied and denote the set of finite poles of the function $R\left(x\right)$ by $\Gamma$.
\begin{description}
\item[\underline{Step 1.}] For each $c\in\Gamma\bigcup\left\{\infty\right\}$ we define a rational function $\left[\sqrt{R}\right]_c$ and two complex numbers $\alpha_{c}^{+}$ and $\alpha_{c}^{-}$ as described below.
\begin{itemize}
\item[$\left(c_1\right)$] If $c\in\Gamma$ is a pole of order 1, then
\begin{equation*}
\left[\sqrt{R}\right]_c=0,\quad \alpha_{c}^{+}=\alpha_{c}^{-}=1.
\end{equation*}
\item[$\left(c_2\right)$] If $c\in\Gamma$ is a pole of order 2, then
\begin{equation*}
\left[\sqrt{R}\right]_c=0.
\end{equation*}

Let $b$ be the coefficient of $\displaystyle\frac{1}{\left(x-c\right)^2}$ in the partial fraction expansion of $R\left(x\right)$. Then
\begin{equation*}
\alpha_{c}^{\pm}=\frac{1}{2}\pm\frac{1}{2}\sqrt{1+4b}.
\end{equation*}
\item[$\left(c_3\right)$] If $c\in\Gamma$ is a pole of order $2\nu\geq 4$ (the order must be even due to the necessary conditions, stated in Theorem~\ref{Theorem2}), then $\left[\sqrt{R}\right]_c$ is the sum of terms involving $\displaystyle\frac{1}{\left(x-c\right)^i}$ for $2\leq i\leq\nu$ in the Laurent expansion of $\sqrt{R}$ at $c$. There are two possibilities for $\left[\sqrt{R}\right]_c$ that differ by sign; we can choose one of them. Thus,
\begin{equation*}
\left[\sqrt{R}\right]_c=\frac{a}{\left(x-c\right)^{\nu}}+\cdots+\frac{d}{\left(x-c\right)^2}.
\end{equation*}

Let $b$ be the coefficient of $\displaystyle\frac{1}{\left(x-c\right)^{\nu+1}}$ in $R-\left[\sqrt{R}\right]^2_c$. Then
\begin{equation*}
\alpha_{c}^{\pm}=\frac{1}{2}\left(\pm\frac{b}{a}+\nu\right).
\end{equation*}
\item[$\left(\infty_1\right)$] If the order of $R\left(x\right)$ at $x=\infty$ is greater than 2, then
\begin{equation*}
\left[\sqrt{R}\right]_{\infty}=0,\quad \alpha_{\infty}^{+}=1,\quad \alpha_{\infty}^{-}=0.
\end{equation*}
\item[$\left(\infty_2\right)$] If the order of $R\left(x\right)$ at $x=\infty$ is 2, then
\begin{equation*}
\left[\sqrt{R}\right]_{\infty}=0.
\end{equation*}

Let $b$ be the coefficient of $\displaystyle\frac{1}{x^2}$ in the Laurent series expansion of $R\left(x\right)$ at $x=\infty$. Then
\begin{equation*}
\alpha_{\infty}^{\pm}=\frac{1}{2}\pm\frac{1}{2}\sqrt{1+4b}.
\end{equation*}
\item[$\left(\infty_3\right)$] If the order $R\left(x\right)$ at $x=\infty$ is $-2\nu\leq 0$ (it is even due to the necessary conditions stated in Theorem~\ref{Theorem2}), then the function $\left[\sqrt{R}\right]_{\infty}$ is the sum of terms involving $x^i$, $0\leq i\leq\nu$ of the Laurent expansion of $\sqrt{R}$ at $x=\infty$ (one of the two possibilities can be chosen). Thus,
\begin{equation*}
\left[\sqrt{R}\right]_{\infty}=ax^{\nu}+\cdots+d.
\end{equation*}

Let $b$ be the coefficient of $x^{\nu-1}$ in $R-\left(\left[\sqrt{R}\right]_{\infty}\right)^2$. Then we have:
\begin{equation*}
\alpha_{\infty}^{\pm}=\frac{1}{2}\left(\pm\frac{b}{a}-\nu\right).
\end{equation*}
\end{itemize}

\item[\underline{Step 2.}] For each family $s=\left(s\left(c\right)\right)_{c\in\Gamma\bigcup\left\{\infty\right\}}$, where $s\left(c\right)$ are either $+$ or $-$ let
\begin{equation}\label{49}
d=\alpha_{\infty}^{s\left(\infty\right)}-\sum\limits_{c\in\Gamma}\alpha_c^{s\left(c\right)}.
\end{equation}

If $d$ is a non -- negative integer, then we introduce the function
\begin{equation}\label{50}
\theta=\sum\limits_{c\in\Gamma}\left(s\left(c\right)\left[\sqrt{R}\right]_{c}+\frac{\alpha_{c}^{s\left(c\right)}}{x-c}\right)
+s\left(\infty\right)\left[\sqrt{R}\right]_{\infty}
\end{equation}

If $d$ is not a non -- negative integer, then the family $s$ should be discarded. If all tuples $s$ have been rejected, then Case~1 cannot hold.

\item[\underline{Step 3.}] For each family $s$ from Step~2, we search for a monic polynomial $P$ of degree $d$ (the constant $d$ is defined by the formula~\eqref{49}), satisfying the differential equation
\begin{equation}\label{51}
P''+2\theta P'+\left(\theta'+\theta^2-R\right)P=0.
\end{equation}

If such a polynomial exists, then
\begin{equation*}
\eta=P\exp\left(\int\theta\left(x\right)dx\right)
\end{equation*}
is the solution of the differential equation~\eqref{48}. If for each tuple $s$ found on Step 2, we cannot find such
a polynomial $P$, then Case 1 cannot hold for the differential equation~\eqref{48}.
\end{description}

Now we state the Kovacic algorithm to search for a solution of type 2 of differential equation~\eqref{48}. We denote the set of finite poles of the function $R\left(x\right)$ by $\Gamma$.
\begin{description}
\item[\underline{Step 1.}] For each $c\in\Gamma\bigcup\left\{\infty\right\}$ we define the set $E_c$ as follows.
\begin{itemize}
\item[$\left(c_1\right)$] If $c\in\Gamma$ is a pole of order 1, then
\begin{equation*}
E_c=\left\{4\right\}.
\end{equation*}
\item[$\left(c_2\right)$] If $c\in\Gamma$ is a pole of order 2 and if $b$ is the coefficient of $\frac{1}{\left(x-c\right)^2}$ in the partial fraction expansion of $R\left(x\right)$, then
\begin{equation*}
E_c=\left\{\left(2+k\sqrt{1+4b}\right)\bigcap\mathbb{Z}\right\},\; k=0, \pm2.
\end{equation*}

\item[$\left(c_3\right)$] If $c\in\Gamma$ is a pole of order $\nu>2$, then
\begin{equation*}
E_c=\left\{\nu\right\}.
\end{equation*}
\item[$\left(\infty_1\right)$] If $R\left(x\right)$ has order $>2$ at $x=\infty$, then
\begin{equation*}
E_{\infty}=\left\{0, 2, 4\right\}.
\end{equation*}

\item[$\left(\infty_2\right)$] If $R\left(x\right)$ has order 2 at $x=\infty$ and $b$ is the coefficient of $\frac{1}{x^2}$ in the Laurent series expansion of $R$ at $x=\infty$, then
\begin{equation*}
E_{\infty}=\left\{\left(2+k\sqrt{1+4b}\right)\bigcap\mathbb{Z}\right\},\; k=0, \pm2.
\end{equation*}

\item[$\left(\infty_3\right)$] If $R\left(x\right)$ has order $\nu<2$ at $x=\infty$, then
\begin{equation*}
E_{\infty}=\left\{\nu\right\}.
\end{equation*}
\end{itemize}

\item[\underline{Step 2.}] Let us consider the families $s=\left(e_\infty, e_c\right)$, $c\in\Gamma$, where $e_c\in E_c$, $e_\infty\in E_\infty$ and at least one of these numbers is odd. Let
\begin{equation}\label{52}
d=\frac{1}{2}\left(e_{\infty}-\sum\limits_{c\in\Gamma} e_c\right).
\end{equation}

If $d$ is a non -- negative integer, the family should be retained, otherwise it should be discarded.

\item[\underline{Step 3.}] For each family retained from Step~2, we form the rational function
\begin{equation}\label{53}
\theta=\frac{1}{2}\sum\limits_{c\in\Gamma}\frac{e_c}{x-c}
\end{equation}
and search for a monic polynomial $P$ of degree $d$ (the constant $d$ is defined by the formula~\eqref{52}) such, that
\begin{equation}\label{54}
P'''+3\theta P''+\left(3\theta^2+3\theta'-4R\right)P'+\left(\theta''+3\theta\theta'+\theta^3-4R\theta-2R'\right)P=0.
\end{equation}

If success is achieved, we set
\begin{equation*}
\varphi=\theta+\frac{P'}{P}
\end{equation*}
and let $\omega$ be a solution of the quadratic equation (algebraic equation of degree 2) of the form:
\begin{equation*}
\omega^2-\varphi\omega+\frac{1}{2}\varphi'+\frac{1}{2}\varphi^2-R=0.
\end{equation*}

Then
\begin{equation*}
\eta=\exp\left(\int\omega\left(x\right)dx\right) -
\end{equation*}
is a solution of the differential equation~\eqref{48}. If success is not achieved, then Case~2 cannot hold for the differential equation~\eqref{48}.
\end{description}

Similarly the Kovacic algorithm is stated to search for a liouvillian solutions of type 3 of the differential equation~\eqref{48}. Let us apply now this algorithm to search liouvillian solutions of the second order linear differential equation~\eqref{45}.

\subsection{Application of the Kovacic algorithm to the differential equation~\eqref{45}. General case.}

So, the differential equation being investigated has the form~\eqref{45}. In this equation we make a substitution according to~\eqref{47} and reduce it to the form~\eqref{48}:
\begin{equation}\label{55}
\frac{d^2y}{dx^2}=R\left(x\right)y.
\end{equation}

Here the function $R\left(x\right)$ takes the form:
\begin{equation}\label{56}
R\left(x\right)=\frac{U\left(x\right)}{V\left(x\right)},
\end{equation}
\begin{equation*}
\begin{array}{l}
U\left(x\right)\!=\!-\left(1+4d_1^2\right)d_1^2x^{16}+8\left(2d_1^2-1\right)d_1^2hx^{14}+4\left(2d_1^2+5\right)d_1k_1x^{13}-
8\left(2d_1^2-7\right)\left(h^2-1\right)d_1^2x^{12}\!-\\ \\
-8\left(4d_1^2+19\right)d_1k_1hx^{11}+8\left(\left(1+17d_1^2-2d_1^4\right)k_1^2-12\left(h^2-1\right)d_1^2h\right)x^{10}
+576\left(h^2-1\right)k_1^3d_1x^3+\\ \\
+16\left(2\left(h^2-1\right)d_1^2+29h^2-5\right)d_1k_1x^9+8\left(6\left(h^2-1\right)^2d_1^2-\left(7+40d_1^2\right)k_1^2h\right)x^8-
240k_1^4hx^2+\\ \\
+32\left(\left(d_1^2-2\right)k_1^2-21\left(h^2-1\right)h\right)d_1k_1x^7+16\left(14\left(h^2-1\right)d_1^2+8h^2-5\right)k_1^2x^6+
288k_1^5d_1x+\\ \\
+96\left(4\left(h^2-1\right)^2-3k_1^2h\right)d_1k_1x^5+4\left(\left(32d_1^2+35\right)k_1^2-24\left(h^2-1\right)h\right)k_1^2x^4+
48\left(h^2-1\right)k_1^4.
\end{array}
\end{equation*}
\begin{equation*}
V\left(x\right)=4\left(d_1x^3-k_1\right)^2\left(x^6-4hx^4+4\left(h^2-1\right)x^2+4k_1^2\right)^2.
\end{equation*}

Thus, it is easy to see, that the function $R\left(x\right)$ has nine finite poles of the second order. Let us denote the roots of the polynomial
\begin{equation}\label{57}
x^6-4hx^4+4\left(h^2-1\right)x^2+4k_1^2=0
\end{equation}
by $x_1$, $x_2$, $x_3$, $x_4$, $x_5$, $x_6$. Note that this polynomial contains only the terms of even degree, therefore its roots satisfy the conditions:
\begin{equation*}
x_2=-x_1,\qquad x_4=-x_3,\qquad x_6=-x_5.
\end{equation*}

Let us denote the roots of the polynomial
\begin{equation}\label{58}
d_1x^3-k_1=0
\end{equation}
by $x_7$, $x_8$, $x_9$. Now let us consider the partial fraction expansion of the function $R\left(x\right)$. It has the form:
\begin{equation*}
R\left(x\right)=-\frac{3}{16}\sum\limits_{i=1}^6\frac{1}{\left(x-x_i\right)^2}+\sum\limits_{i=1}^9\frac{\gamma_i\left(x_i\right)}
{x-x_i}+\frac{3}{4}\sum\limits_{i=7}^9\frac{1}{\left(x-x_i\right)^2}.
\end{equation*}

The coefficients $\gamma_i\left(x_i\right)$, $i=1, 2,\ldots 9$ have a very complicated form and we do not write them explicitly here. It is possible to note the following properties of the partial fraction expansion of the function $R\left(x\right)$.
\begin{enumerate}
\item The coefficients $b_1,\ldots, b_6$ of $\displaystyle\frac{b_i}{\left(x-x_i\right)^2}$, $i=1,\ldots, 6$ are all equal
\begin{equation*}
b_i=-\frac{3}{16}, \quad i=1,\ldots, 6.
\end{equation*}
\item The coefficients $b_7$, $b_8$, $b_9$ of $\displaystyle\frac{b_i}{\left(x-x_i\right)^2}$, $i=7, 8, 9$ are all equal
\begin{equation*}
b_i=\frac{3}{4}, \quad i=7, 8, 9.
\end{equation*}
\item The Laurent expansion of $R\left(x\right)$ at $x=\infty$ has the form:
\begin{equation*}
R\left(x\right)=-\frac{\left(1+4d_1^2\right)}{4x^2}+O\left(\frac{1}{x^4}\right)
\end{equation*}
\end{enumerate}

Thus, we have
\begin{equation*}
b_{\infty}=-\frac{1}{4}-d_1^2,
\end{equation*}
and therefore
\begin{equation*}
1+4b_{\infty}=-4d_1^2
\end{equation*}

This means that the numbers $\alpha^{\pm}_{\infty}$ calculating during the application of the Kovacic algorithm for searching the liouvillian solutions of type 1, are complex numbers if $d_1\ne 0$. All the remaining numbers $\alpha^{\pm}_c$ are rational. They are presented in the following Table. Therefore, the number $d$, calculated by formula~\eqref{49} in the process of searching for liouvillian solutions of type 1, is a complex number for $d_1\ne 0$. This fact indicates the absence of liouvillian solutions of type 1 for $d_1\ne 0$.
\begin{center}
\begin{tabular}{ |c|c|c|c|c|c|c|c|c|c| }
 \hline
  & $x_1$ & $x_2$ & $x_3$ & $x_4$ & $x_5$ & $x_6$ & $x_7$ & $x_8$ & $x_9$ \\ \hline
  \rule{0pt}{4ex} $\alpha_c^+$ & $\displaystyle\frac{3}{4}$ & $\displaystyle\frac{3}{4}$ & $\displaystyle\frac{3}{4}$ & $\displaystyle\frac{3}{4}$ & $\displaystyle\frac{3}{4}$ & $\displaystyle\frac{3}{4}$ & $\displaystyle\frac{3}{2}$ & $\displaystyle\frac{3}{2}$ & $\displaystyle\frac{3}{2}$\\[2ex] \hline
 \rule{0pt}{4ex} $\alpha_c^-$ & $\displaystyle\frac{1}{4}$ & $\displaystyle\frac{1}{4}$ & $\displaystyle\frac{1}{4}$ & $\displaystyle\frac{1}{4}$ & $\displaystyle\frac{1}{4}$ & $\displaystyle\frac{1}{4}$ & $-\displaystyle\frac{1}{2}$ & $-\displaystyle\frac{1}{2}$ & $-\displaystyle\frac{1}{2}$\\[2ex] \hline
\end{tabular}
{\vskip 0.2cm}
Table. Numbers $\alpha^{\pm}_c$ for searching liouvillian solutions of type 1.
\end{center}

Moreover, the coefficient $b_{\infty}$ coincides with the number $\gamma$ calculating during the checking of the necessary conditions of existence of liouvillian solutions of type 3 for the differential equation~\eqref{48}. According to this necessary conditions for existence of liouvillian solutions of type 3 the number
\begin{equation*}
\sqrt{1+4\gamma}=\sqrt{1+4b_{\infty}}
\end{equation*}
should be rational. However, when $d_1\ne 0$ this number is pure imaginary. Thus, we can state that for $d_1\ne 0$ the second order linear differential equation~\eqref{55} (or~\eqref{45}) do not have liouvillian solutions of type 3. Thus, the following Theorem is valid.
\begin{thm}\label{Theorem3}
If all roots of polynomials~\eqref{57} and~\eqref{58} are distinct and $d_1\ne 0$, then the problem of motion of a heavy rigid body with a fixed point in the Hess case has not liouvillian solutions of type 1 and type 3.
\end{thm}

According to Theorem~\ref{Theorem3}, equation~\eqref{55} can have liouvillian solutions of type 1 only when $d_1=0$, i.e. when the moving rigid body with a fixed point is the Lagrange top. To search liouvillian solutions of type 1 for the differential equation~\eqref{55} in the Lagrange integrable case we put $d_1=0$ in this equation.

\subsection{Investigation of the Lagrange case.}

When $d_1=0$ we can write equation~\eqref{55} as follows:
\begin{equation}\label{59}
\frac{d^2y}{dx^2}=R\left(x\right)y=\frac{U\left(x\right)}{V\left(x\right)}y,
\end{equation}
\begin{equation*}
\begin{array}{l}
U\left(x\right)=2x^{10}-14hx^8+4\left(8h^2-5\right)x^6+\left(35k_1^2-24h\left(h^2-1\right)\right)x^4-60k_1^2hx^2+12k_1^2
\left(h^2-1\right),
\end{array}
\end{equation*}
\begin{equation*}
V_1\left(x\right)=\left(x^6-4hx^4+4\left(h^2-1\right)x^2+4k_1^2\right)^2.
\end{equation*}

We assume that in the equation~\eqref{59} there should be $k_1\ne 0$. Otherwise, the conditions~\eqref{44} are satisfied and it is impossible to obtain from the equation~\eqref{41} the second-order linear differential equation~\eqref{45} (or~\eqref{55}), from which for $d_1=0$ we obtain~\eqref{59}.

It is easy to see that the poles of the function $R\left(x\right)$ are the roots of the polynomial~\eqref{57}. We assume that the polynomial~\eqref{57} has no multiple roots (the possibility of multiple roots for the polynomial~\eqref{57} is considered below, in Section 12). Then the function $R\left(x\right)$ has six finite poles of the second order. We denote these poles by $x_1$, $x_2$, $x_3$, $x_4$, $x_5$, $x_6$. Now let us consider the partial fraction expansion of the function $R\left(x\right)$. It has the form:
\begin{equation*}
R\left(x\right)=-\frac{3}{16}\sum\limits_{i=1}^6\frac{1}{\left(x-x_i\right)^2}+\sum\limits_{i=1}^6\frac{\gamma_i\left(x_i\right)}
{x-x_i},
\end{equation*}
\begin{equation*}
\begin{array}{l}
\gamma_i\left(x_i\right)=\displaystyle\frac{\left(48\left(h^2-1\right)^3-24k_1^2h\left(h^2-1\right)\left(h^2-13\right)-
\left(121h^2-177\right)k_1^4\right)x_i}{16k_1^2\left(27k_1^4+8k_1^2h\left(h^2-9\right)-16\left(h^2-1\right)^2\right)}+\\ \\
+\displaystyle\frac{\left(32k_1^2\left(h^4-1\right)-48h\left(h^2-1\right)^2+3k_1^2h\left(61k_1^2-128h\right)\right)x_i^3}
{16k_1^2\left(27k_1^4+8k_1^2h\left(h^2-9\right)-16\left(h^2-1\right)^2\right)}+\\ \\
+\displaystyle\frac{\left(48\left(h^2-1\right)^2-40k_1^2h\left(h^2-1\right)-3\left(75k_1^2-128h\right)k_1^2\right)x_i^5}
{64k_1^2\left(27k_1^4+8k_1^2h\left(h^2-9\right)-16\left(h^2-1\right)^2\right)}.
\end{array}
\end{equation*}

It is possible to note the following properties of the partial fraction expansion of the function $R\left(x\right)$.
\begin{enumerate}
\item The coefficients $b_1,\ldots, b_6$ of $\displaystyle\frac{b_i}{\left(x-x_i\right)^2}$, $i=1,\ldots, 6$ are all equal
\begin{equation*}
b_i=-\frac{3}{16}, \quad i=1,\ldots, 6.
\end{equation*}
\item The Laurent expansion of $R\left(x\right)$ at $x=\infty$ has the form:
\begin{equation*}
R\left(x\right)=\frac{2}{x^2}+O\left(\frac{1}{x^4}\right).
\end{equation*}
\end{enumerate}

Our goal is to find liouvillian solutions of type 1, type 2 and type 3 of the differential equation~\eqref{59}. To find liouvillian solutions of~\eqref{59} we will use the Kovacic algorithm. First we search liouvillian solutions of type 1. According to the Kovacic algorithm, we calculate the numbers $\alpha^{\pm}_c$. For all the roots $x=x_i$, $i=1,\ldots, 6$ these numbers are equal \begin{equation}\label{60}
\alpha_{x_i}^{+}=\frac{3}{4},\qquad \alpha_{x_i}^{-}=\frac{1}{4}, \qquad i=1,\ldots 6.
\end{equation}

The numbers $\alpha^{\pm}_{\infty}$ are equal
\begin{equation}\label{61}
\alpha^{+}_{\infty}=2,\qquad \alpha^{-}_{\infty}=-1.
\end{equation}

Note that the polynomial~\eqref{57} contains the terms of even degree, therefore its roots can be represented as follows:
\begin{equation*}
x_1=y_1,\quad x_2=-x_1=-y_1,\quad x_3=y_2,\quad x_4=-x_3=-y_2,\quad x_5=iy_3,\quad x_6=-x_5=-iy_3.
\end{equation*}

Since the numbers $\alpha^{\pm}_{x_i}$ are determined by~\eqref{60} and the numbers $\alpha^{\pm}_{\infty}$ are determined by~\eqref{61} therefore the number $d$ calculating according to~\eqref{49} can be only zero. This takes place for the following sets of signs $+\,$ and $-:\,$ $s=\left(s\left(\infty\right),\,s\left(x_1\right),\,s\left(x_2\right),\,s\left(x_3\right),\,s\left(x_4\right),\,s\left(x_5\right),\,
s\left(x_6\right)\right)$
\begin{equation*}
\begin{array}{l}
s_1=\left(+,\,+,\,-,\,-,\,-,\,-,\,-\right),\quad s_2=\left(+,\,-,\,+,\,-,\,-,\,-,\,-\right), \quad
s_3=\left(+,\,-,\,-,\,+,\,-,\,-,\,-\right), \\ \\
s_4=\left(+,\,-,\,-,\,-,\,+,\,-,\,-\right),\quad s_5=\left(+,\,-,\,-,\,-,\,-,\,+,\,-\right),\quad
s_6=\left(+,\,-,\,-,\,-,\,-,\,-,\,+\right).
\end{array}
\end{equation*}

We must check all these sets. Let us check now the set $s_1$. For this set of signs let us find the function $\theta$ according to the formula~\eqref{50}. This function has the form:
\begin{equation*}
\theta=\frac{3}{4\left(x-y_1\right)}+\frac{1}{4\left(x+y_1\right)}+\frac{1}{4\left(x-y_2\right)}+\frac{1}{4\left(x+y_2\right)}+
\frac{1}{4\left(x-iy_3\right)}+\frac{1}{4\left(x+iy_3\right)}.
\end{equation*}

The polynomial $P$ of degree $d=0$ has $P\equiv 1$. The substitution of the polynomial $P$ and the function $\theta$ with the differential equation~\eqref{51} reduces it to the form:
\begin{equation}\label{62}
\frac{d\theta}{dx}+\theta^2-R\left(x\right)=0.
\end{equation}

The conditions on the parameters $k_1$ and $h$ for which the left hand side of~\eqref{62} becomes zero are the conditions of existence of liouvillian solutions of type 1 for the differential equation~\eqref{59}. Before we substitute the function $\theta$ to the equation~\eqref{62}, let us simplify it. We can rewrite this function in the form:
\begin{equation*}
\theta=\frac{1}{2\left(x-y_1\right)}+\frac{1}{4}\left[\frac{1}{x-y_1}+\frac{1}{x+y_1}+\frac{1}{x-y_2}+\frac{1}{x+y_2}+
\frac{1}{x-iy_3}+\frac{1}{x+iy_3}\right].
\end{equation*}

Expression in square brackets can be represented as follows:
\begin{equation*}
\frac{1}{x-y_1}+\frac{1}{x+y_1}+\frac{1}{x-y_2}+\frac{1}{x+y_2}+
\frac{1}{x-iy_3}+\frac{1}{x+iy_3}=\frac{2x\left(3x^4-8hx^2+4\left(h^2-1\right)\right)}{x^6-4hx^4+4\left(h^2-1\right)x^2+4k_1^2}.
\end{equation*}

Thus, function $\theta$ has the form:
\begin{equation*}
\theta=\frac{1}{2\left(x-y_1\right)}+\frac{x\left(3x^4-8hx^2+4\left(h^2-1\right)\right)}{2\left(x^6-4hx^4+4\left(h^2-1\right)x^2
+4k_1^2\right)}.
\end{equation*}

If we substitute this function $\theta$ to~\eqref{62} we obtain in the left hand side of~\eqref{62} the rational expression. The numerator of this expression is the fifth degree polynomial of the form:
\begin{equation*}
P_5\left(x\right)=4y_1x^5+\cdots
\end{equation*}

The necessary condition for this polynomial to vanish is the condition $y_1=0$, that is, the polynomial~\eqref {57} must have a zero root. This condition is equivalent to the condition $k_1=0$. However, we previously suggested that $k_1\ne 0$. Therefore, $y_1\ne 0$ and we can state that the expression in the left side of the equation~\eqref{62} does not vanish. Thus, for the set $s_1$ there are no liouvillian solutions of type 1 in the problem of the motion of a heavy rigid body with a fixed point in the Hess case.

Similarly, the fact of the absence of liouvillian solutions of type 1 for the sets $s_2, \ldots, s_6$ is established. So, we can state that the following theorem is true.

\begin{thm}\label{Theorem4}
If all roots of polynomial~\eqref{57} are distinct and $d_1=0$, $k_1\ne 0$, then the problem of motion of a heavy rigid body with a fixed point in the Hess case has not liouvillian solutions of type $1$.
\end{thm}

Now we continue the study of the existence of liouvillian solutions in the problem of the motion of the Hess top under the additional condition $d_1=0$, that is, in the problem of the motion of the Lagrange top when the corresponding second-order linear differential equation has the form~\eqref{59}. Let us study here the problem of the existence of liouvillian solutions of type 2 for the differential equation~\eqref{59}.

According to the Kovacic algorithm for searching the liouvillian solutions of type 2 we should find the sets $E_{x_i}$, $i=1,\ldots, 6$ which correspond to finite poles of $R\left(x\right)$ and the set $E_{\infty}$ which corresponds to the pole of $R\left(x\right)$ at $x=\infty$. For the finite poles $x=x_i$, $i=1,\ldots, 6$, which are the roots of the polynomial~\eqref{57}, the sets $E_{x_i}$ are all equal
\begin{equation*}
E_{x_i}=\{1, 2, 3\}, \quad i=1,\ldots, 6.
\end{equation*}

The set $E_{\infty}$ has the following form:
\begin{equation*}
E_{\infty}=\{-4, 2, 8\}.
\end{equation*}

Now we should calculate the constant $d$ using~\eqref{52}. Note that the minimal value of the sum
\begin{equation*}
\sum\limits_{x_i\in\Gamma} e_{x_i}
\end{equation*}
equals 6. Therefore, the maximal value of $d$, calculated according to~\eqref{52}, equals $d=1$. The family $s=\left(e_{\infty}, e_1, e_2, e_3, e_4, e_5, e_6\right)$ corresponding to $d=1$ is
\begin{equation*}
s_1=\left(8, 1, 1, 1, 1, 1, 1\right).
\end{equation*}

We also have several families, for which we have $d=0$. The following families correspond to $d=0$:
\begin{equation*}
\begin{array}{l}
s_2=\left(8, 3, 1, 1, 1, 1, 1\right),\quad s_3=\left(8, 1, 3, 1, 1, 1, 1\right),\quad s_4=\left(8, 1, 1, 3, 1, 1, 1\right),\\ \\
s_5=\left(8, 1, 1, 1, 3, 1, 1\right),\quad s_6=\left(8, 1, 1, 1, 1, 3, 1\right),\quad s_7=\left(8, 1, 1, 1, 1, 1, 3\right),\\ \\
s_8=\left(8, 2, 2, 1, 1, 1, 1\right),\quad s_9=\left(8, 2, 1, 2, 1, 1, 1\right),\quad s_{10}=\left(8, 2, 1, 1, 2, 1, 1\right), \\ \\
s_{11}=\left(8, 2, 1, 1, 1, 2, 1\right),\quad s_{12}=\left(8, 2, 1, 1, 1, 1, 2\right),\quad s_{13}=\left(8, 1, 2, 2, 1, 1, 1\right),\\ \\
s_{14}=\left(8, 1, 2, 1, 2, 1, 1\right),\quad s_{15}=\left(8, 1, 2, 1, 1, 2, 1\right),\quad s_{16}=\left(8, 1, 2, 1, 1, 1, 2\right),\\ \\
s_{17}=\left(8, 1, 1, 2, 2, 1, 1\right),\quad s_{18}=\left(8, 1, 1, 2, 1, 2, 1\right),\quad s_{19}=\left(8, 1, 1, 2, 1, 1, 2\right),\\ \\
s_{20}=\left(8, 1, 1, 1, 2, 2, 1\right),\quad s_{21}=\left(8, 1, 1, 1, 2, 1, 2\right),\quad s_{22}=\left(8, 1, 1, 1, 1, 2, 2\right).
\end{array}
\end{equation*}

We must check all these families. We start with the family $s_1$. According to the algorithm let us find the function $\theta$ by the formula~\eqref{53}. Since the coefficients at all finite poles are the same, we can write the function $\theta$ in explicit form. For the family $s_1$ this function has the form:
\begin{equation*}
\theta=\frac{3x^5-8hx^3+4\left(h^2-1\right)x}{x^6-4hx^4+4\left(h^2-1\right)x^2+4k_1^2}.
\end{equation*}

The polynomial $P$ of degree $d=1$ $\left(P\equiv x+b\right)$ should identically satisfy differential equation~\eqref{54}. After substitution of the polynomial $P$ and the functions $\theta$ and $R\left(x\right)$ in~\eqref{54}, we obtain in the left hand side of it the following expression:
\begin{equation*}
\frac{2Bx\left(6h-5x^2\right)}{x^6-4hx^4+4\left(h^2-1\right)x^2+4k_1^2}.
\end{equation*}

This expression becomes identically zero when $B=0$. Therefore, the polynomial $P\equiv x$ exists for all values $h$ and
$k_1$. Thus, we can state the following theorem.

\begin{thm}\label{Theorem5}
In the Lagrange integrable case of motion $d_1=0$ under the Hess conditions~\eqref{7} and also under condition that all the roots of polynomial~\eqref{57} are distinct and $k_1\ne 0$, all solutions of a linear differential equation~\eqref{55} are liouvillian solutions of type $2$.
\end{thm}

Indeed, it is easy to find the general solution of equation~\eqref{45} for $d_1=0$. In this case the differential equation~\eqref{45} takes the form
\begin{equation*}
\frac{d^2u}{dx^2}+\frac{4\left(x^6-3hx^4+2\left(h^2-1\right)x^2+k_1^2\right)}{x\left(x^6-4hx^4+4\left(h^2-1\right)x^2+4k_1^2\right)}
\frac{du}{dx}-\frac{k_1^2u}{x^2\left(x^6-4hx^4+4\left(h^2-1\right)x^2+4k_1^2\right)}=0,
\end{equation*}
and its general solution can be written as follows:
\begin{equation*}
\begin{array}{l}
u\left(x\right)={\rm C}_1\exp\left(\displaystyle\int\displaystyle\frac{k_1dx}{x\sqrt{x^6-4hx^4+4\left(h^2-1\right)x^2+4k_1^2}}\right)+\\ \\
+{\rm C}_2\exp\left(-\displaystyle\int\displaystyle\frac{k_1dx}{x\sqrt{x^6-4hx^4+4\left(h^2-1\right)x^2+4k_1^2}}\right).
\end{array}
\end{equation*}

If we check the other families $s_2, \ldots, s_{22}$ we obtain the same results on the existing of liouvillian solutions of type 2 for the differential equation~\eqref{59}.

Now we come back to the investigation of the general case $d_1\ne 0$ and consider the problem of existence of liouvillian solutions of type 2 for the differential equation~\eqref{55}.

\subsection{Existence of liouvillian solutions of type 2 in general case.}

So we are going to study the problem of existence of liouvillian solutions of type 2 for the differential equation~\eqref{45} (or~\eqref{55}). According to the Kovacic algorithm we firstly define the sets $E_c$ and $E_{\infty}$ for every pole of the function $R\left(x\right)$. For the finite poles $x=x_i$, $i=1,\ldots, 6$, which are roots of the polynomial~\eqref{57}, these sets $E_{x_i}$ have the form:
\begin{equation*}
E_{x_i}=\{1, 2, 3\}, \quad i=1,\ldots, 6.
\end{equation*}

For the finite poles $x=x_i$, $i=7, 8, 9$, which are roots of the polynomial~\eqref{58}, these sets $E_{x_i}$ have the form:
\begin{equation*}
E_{x_i}=\{-2, 2, 6\}, \quad i=7, 8, 9.
\end{equation*}

The set $E_{\infty}$ contains only one element and this set equals
\begin{equation*}
E_{\infty}=\{2\}.
\end{equation*}

Now we should calculate the constant $d$ using the formula~\eqref{52}. Note that the minimal values of the sum of the elements of sets corresponding to finite poles is zero. Therefore the maximal value of $d$, calculated according to~\eqref{52}, equals $d=1$. The value $d=1$ corresponds to the set $s=\left(e_{\infty}, e_1, e_2, e_3, e_4, e_5, e_6, e_7, e_8, e_9\right)$, in which the elements $e_{\infty} $ and $e_i$, $i=1, 2, \ldots 9$ are equal
\begin{equation*}
s_1=\left(2, 1, 1, 1, 1, 1, 1, -2, -2, -2\right).
\end{equation*}

We also have several families for which the constant $d$, calculated by the formula~\eqref{52}, equals to zero. The following families correspond to $d=0$:
\begin{equation*}
\begin{array}{l}
s_2=\left(2, 3, 1, 1, 1, 1, 1, -2, -2, -2\right),\quad s_3=\left(2, 1, 3, 1, 1, 1, 1, -2, -2, -2\right),\\ \\
s_4=\left(2, 1, 1, 3, 1, 1, 1, -2, -2, -2\right),\quad s_5=\left(2, 1, 1, 1, 3, 1, 1, -2, -2, -2\right),\\ \\
s_6=\left(2, 1, 1, 1, 1, 3, 1, -2, -2, -2\right),\quad s_7=\left(2, 1, 1, 1, 1, 1, 3, -2, -2, -2\right),\\ \\
s_8=\left(2, 2, 2, 1, 1, 1, 1, -2, -2, -2\right),\quad s_9=\left(2, 2, 1, 2, 1, 1, 1, -2, -2, -2\right),\\ \\
s_{10}=\left(2, 2, 1, 1, 2, 1, 1, -2, -2, -2\right),\quad s_{11}=\left(2, 2, 1, 1, 1, 2, 1, -2, -2, -2\right),\\ \\
s_{12}=\left(2, 2, 1, 1, 1, 1, 2, -2, -2, -2\right),\quad s_{13}=\left(2, 1, 2, 2, 1, 1, 1, -2, -2, -2\right),\\ \\
s_{14}=\left(2, 1, 2, 1, 2, 1, 1, -2, -2, -2\right),\quad s_{15}=\left(2, 1, 2, 1, 1, 2, 1, -2, -2, -2\right),\\ \\
s_{16}=\left(2, 1, 2, 1, 1, 1, 2, -2, -2, -2\right),\quad s_{17}=\left(2, 1, 1, 2, 2, 1, 1, -2, -2, -2\right),
\end{array}
\end{equation*}
\begin{equation*}
\begin{array}{l}
s_{18}=\left(2, 1, 1, 2, 1, 2, 1, -2, -2, -2\right),\quad s_{19}=\left(2, 1, 1, 2, 1, 1, 2, -2, -2, -2\right),\\ \\
s_{20}=\left(2, 1, 1, 1, 2, 2, 1, -2, -2, -2\right),\quad s_{21}=\left(2, 1, 1, 1, 2, 1, 2, -2, -2, -2\right),\\ \\
s_{22}=\left(2, 1, 1, 1, 1, 2, 2, -2, -2, -2\right).
\end{array}
\end{equation*}

We must check all these families. We start with the family $s_1$. According to the algorithm, let us find the function $\theta$ by the formula~\eqref{53}. Since the coefficients at all finite poles $x=x_i$, $i=1, 2, \ldots, 6$ are the same, and the coefficients at all poles $x=x_i$, $i=7, 8, 9$ are the same, then we can write the function $\theta$ in explicit form. For the set $s_1$ this function has the form
\begin{equation*}
\theta=\frac{3x^5-8hx^3+4\left(h^2-1\right)x}{x^6-4hx^4+4\left(h^2-1\right)x^2+4k_1^2}-\frac{3d_1x^2}{d_1x^3-k_1}.
\end{equation*}

The polynomial $P$ of degree $d=1$
\begin{equation*}
P=x+B
\end{equation*}
should identically satisfy the differential equation~\eqref{54}. After substitution of $P=x+B$ and the functions $\theta$ and $R\left(x\right)$ to the equation~\eqref{54}, we obtain in the left hand side of~\eqref{54} the rational expression. The numerator of this expression has a form of the ninth degree polynomial:
\begin{equation*}
P_9=-Bd_1^2\left(1+4d_1^2\right)x^9+\cdots
\end{equation*}

Let us put $B=0$. Then the numerator of the rational expression in the left hand side of~\eqref{54} takes the form:
\begin{equation*}
P_7=-12xk_1d_1\left(d_1x^3-k_1\right)^2+\cdots
\end{equation*}

This polynomial becomes zero when $k_1=0$ or when $d_1=0$. Therefore we can state the following Theorem based on the verification of the family $s_1$.

\begin{thm}\label{Theorem6}
For the existence of liouvillian solutions of type 2 in the problem of motion of a heavy rigid body with a fixed point in the Hess case one of the two conditions must be satisfied:
\begin{equation*}
d_1=0\quad\mbox{or}\quad k_1=0.
\end{equation*}

In other words, liouvillian solutions of type $2$ can exist either in the case, when the moving rigid body is the Lagrange top, or in the Hess case, if the constant of the area integral is zero.
\end{thm}

The fact, that the problem of motion of a heavy rigid body with a fixed point in the Hess case for $k_1=0$ is integrable in elliptic functions (which are liouvillian functions) was firstly discovered by P.~A.~Nekrasov~\cite{Nekrasov1, Nekrasov2}.

To prove Theorem~\ref{Theorem6} we need to check the families $s_2,\ldots, s_{22}$, and to consider various critical cases, which we will discuss further, in Sections 11 -- 15. We start with families $s_2,\ldots, s_7$. For the family $s_2$ the function $\theta$, calculated by the formula~\eqref{53}, has the form:
\begin{equation*}
\theta=\frac{1}{2}\left(\frac{3}{x-y_1}+\frac{1}{x+y_1}+\frac{1}{x-y_2}+\frac{1}{x+y_2}+\frac{1}{x-iy_3}+\frac{1}{x+iy_3}\right)-
\frac{3d_1x^2}{d_1x^3-k_1}.
\end{equation*}

We slightly simplify this expression for the function $\theta$. We can rewrite it in the form:
\begin{equation*}
\theta=\frac{1}{2}\left[\frac{1}{x-y_1}+\frac{1}{x+y_1}+\frac{1}{x-y_2}+\frac{1}{x+y_2}+\frac{1}{x-iy_3}+\frac{1}{x+iy_3}\right]+
\frac{1}{x-y_1}-\frac{3d_1x^2}{d_1x^3-k_1}.
\end{equation*}

Expression in square brackets can be represented as follows:
\begin{equation*}
\frac{1}{x-y_1}+\frac{1}{x+y_1}+\frac{1}{x-y_2}+\frac{1}{x+y_2}+\frac{1}{x-iy_3}+\frac{1}{x+iy_3}=\frac{2x\left(3x^4-8hx^2+4\left(h^2-1
\right)\right)}{x^6-4hx^4+4\left(h^2-1\right)x^2+4k_1^2}.
\end{equation*}

Finally we have the following expression for the function $\theta$:
\begin{equation*}
\theta=\frac{3x^5-8hx^3+4\left(h^2-1\right)x}{x^6-4hx^4+4\left(h^2-1\right)x^2+4k_1^2}+\frac{1}{x-y_1}-\frac{3d_1x^2}{d_1x^3-k_1}.
\end{equation*}

The polynomial $P$ of degree $d=0$ has the form  $P\equiv 1$. This polynomial should identically satisfy the differential equation~\eqref{54}. After substitution of $P=1$ and the functions $\theta$ and $R\left(x\right)$ in this equation, we obtain in the left hand side of~\eqref{54} the rational expression. The numerator of this expression has a form of the ninth degree polynomial of $x$, the leading coefficient of which is equal to
\begin{equation*}
P_{9}=y_1d_1^2\left(1+4d_1^2\right)x^9+\cdots
\end{equation*}

If we put $y_1=0$ (this condition is equivalent to the condition $k_1=0$), then this polynomial $P_9$ becomes zero. Thus, checking of the family $s_2$ gives the same conditions as the checking of the family $s_1$. These conditions have been formulated in Theorem~\ref{Theorem6}. Similarly we can check the other families $s_3, \ldots, s_7$.

Let us check now the families $s_8,\ldots, s_{22}$. For the family $s_8$ the function $\theta$, calculated by the formula~\eqref{53}, has the form:
\begin{equation*}
\theta=\frac{1}{x-y_1}+\frac{1}{x+y_1}+\frac{1}{2}\left(\frac{1}{x-y_2}+\frac{1}{x+y_2}+\frac{1}{x-iy_3}+\frac{1}{x+iy_3}\right)-
\frac{3d_1x^2}{d_1x^3-k_1}.
\end{equation*}

We rewrite this function as follows:
\begin{equation*}
\theta=\frac{1}{2\left(x\!-\!y_1\right)}+\frac{1}{2\left(x\!+\!y_1\right)}+\frac{1}{2}\left(\frac{1}{x\!-\!y_1}+\frac{1}{x\!+\!y_1}
+\frac{1}{x\!-\!y_2}+\frac{1}{x\!+\!y_2}+\frac{1}{x\!-\!iy_3}+\frac{1}{x\!+\!iy_3}\right)-\frac{3d_1x^2}{d_1x^3\!-\!k_1}.
\end{equation*}

The latter expression can be transformed to the form:
\begin{equation*}
\theta=\frac{3x^5-8hx^3+4\left(h^2-1\right)x}{x^6-4hx^4+4\left(h^2-1\right)x^2+4k_1^2}-\frac{3d_1x^2}{d_1x^3-k_1}+
\frac{x}{x^2-y_1^2}.
\end{equation*}

The polynomial $P$ of degree $d=0$ has the form $P\equiv 1$. This polynomial should identically satisfy the differential equation~\eqref{54}. After substituting $P=1$ and the functions $\theta$ and $R\left(x\right)$ in this equation, we obtain in the left hand side of~\eqref{54} the rational expression. The numerator of this expression has a form of 13th degree polynomial
\begin{equation*}
P_{13}=4y_1^2d_1^2\left(1+d_1^2\right)x^{13}+\cdots
\end{equation*}

If we put $y_1=0$ and $k_1=0$ then this polynomial becomes zero. Thus, checking of the family $s_8$ gives the same conditions as checking of the family $s_1$. These conditions have been formulated in Theorem~\ref{Theorem6}. Similarly we can check the other families $s_9, \ldots, s_{22}$. Checking of these families gives us the same conditions as checking of the family $s_1$.

Finally we can state that the conditions of existence of liouvillian solutions of type 2 for the differential equation~\eqref{55} are formulated in Theorem~\ref{Theorem6}. To confirm the obtained results let us consider the equation~\eqref{55} in the case when $k_1=0$.

\subsection{Existence of liouvillian solutions of type 2 in the case $k_1=0$.}

In the case $k_1=0$, the function $R\left(x\right)$, defined by~\eqref{56}, has the form:
\begin{equation*}
R\left(x\right)=\frac{U\left(x\right)}{V\left(x\right)},
\end{equation*}
\begin{equation*}
U\left(x\right)=-\left(1+4d_1^2\right)x^{8}+8h\left(2d_1^2-1\right)x^{6}+8\left(7-2d_1^2\right)\left(h^2-1\right)x^4-
96h\left(h^2-1\right)x^2+48\left(h^2-1\right)^2,
\end{equation*}
\begin{equation*}
V\left(x\right)=4x^2\left(x^2-2h-2\right)^2\left(x^2-2h+2\right)^2.
\end{equation*}

Thus, the function $R\left(x\right)$ has five finite poles of the second order. One of these finite poles is $x_0=0$ and the other four $x_1$, $x_2$, $x_3$, $x_4$ are the roots of two polynomials:
\begin{equation*}
x^2-2h-2=0
\end{equation*}
(we denote its roots by $x_1$, $x_2$) and
\begin{equation*}
x^2-2h+2=0
\end{equation*}
(we denote its roots by $x_3$, $x_4$). Partial fraction expansion of the function $R\left(x\right)$ has the form:
\begin{equation*}
R\left(x\right)=\frac{3}{4x^2}-\sum\limits_{i=1}^4\frac{3}{16\left(x-x_i\right)^2}-\sum\limits_{i=1}^2\frac{\left(4d_1^2\left(h+1\right)
-2h+1\right)x_i}{32\left(h+1\right)\left(x-x_i\right)}+\sum\limits_{i=3}^4\frac{\left(4d_1^2\left(h-1\right)
-2h-1\right)x_i}{32\left(h-1\right)\left(x-x_i\right)}.
\end{equation*}

It is possible to note the following properties on the partial fraction expansion of the function $R\left(x\right)$.
\begin{enumerate}
\item The coefficient $b_0$ of  $\displaystyle\frac{b_0}{x^2}$ equals
\begin{equation*}
b_0=\frac{3}{4}.
\end{equation*}
\item The coefficients $b_1,\ldots, b_4$ of $\displaystyle\frac{b_i}{\left(x-x_i\right)^2}$, $i=1, 2, 3, 4$ are all equal
\begin{equation*}
b_i=-\frac{3}{16}, \quad i=1, 2, 3, 4.
\end{equation*}
\item The Laurent expansion of $R\left(x\right)$ at $x=\infty$ has the form:
\begin{equation*}
R\left(x\right)=-\frac{\left(1+4d_1^2\right)}{4x^2}+O\left(\frac{1}{x^4}\right).
\end{equation*}
\end{enumerate}

To find liouvillian solutions of type 2 of the differential equation~\eqref{55} in the case $k_1=0$ we will use the Kovacic algorithm. According to the algorithm, let us define the sets $E_{x_i}$, corresponding to finite poles of the function $R\left(x\right)$ and the set $E_{\infty}$ corresponding to the pole of $R\left(x\right)$ at $x=\infty$. For the pole $x=0$ the corresponding set has the form
\begin{equation*}
E_0=\{-2, 2, 6\}.
\end{equation*}

For the finite poles $x_i$, $i=1, 2, 3, 4$ the sets $E_{x_i}$ have the form:
\begin{equation*}
E_{x_i}=\{1, 2, 3\},\quad i=1, 2, 3, 4.
\end{equation*}

The set $E_{\infty}$ contains only one element and this set equals
\begin{equation*}
E_{\infty}=\{2\}.
\end{equation*}

It is easy to see, that the constant $d$, calculated by the formula~\eqref{52} is a non-negative integer only for the family
\begin{equation*}
s_1=\{2, -2, 1, 1, 1, 1\},
\end{equation*}
for which we have $d=0$. Using this family, let us find the function $\theta$ by the formula~\eqref{53}. This function has a form
\begin{equation*}
\theta=-\frac{1}{x}+\frac{2\left(x^2-2h\right)x}{\left(x^2-2h-2\right)\left(x^2-2h+2\right)}.
\end{equation*}

The polynomial $P$ of degree $d=0$ has the form $P\equiv 1$. This polynomial should identically satisfy the differential equation~\eqref{54}. After substitution of $P=1$ and the functions $\theta$ and $R\left(x\right)$ to the differential equation~\eqref{54} we find that this equation is satisfied identically for any values of parameters $d_1$ and $h$. Thus, we can state the following theorem.
\begin{thm}\label{Theorem7}
In the case of motion of a heavy rigid body with a fixed point in the Hess case $\left(d_1\ne 0\right)$, equation~\eqref{55} has liouvillian solutions only in the case $k_1=0$.
\end{thm}

Indeed, when $k_1=0$ the differential equation~\eqref{45} takes the form:
\begin{equation}\label{63}
\frac{d^2u}{dx^2}+\frac{x^4-4\left(h^2-1\right)}{x\left(x^2-2h-2\right)\left(x^2-2h+2\right)}
\frac{du}{dx}+\frac{d_1^2x^2}{\left(x^2-2h-2\right)\left(x^2-2h+2\right)}u=0.
\end{equation}

The general solution of the equation~\eqref{63} has the form:
\begin{equation*}
\begin{array}{l}
u\left(x\right)={\rm C}_1\exp\left(id_1\displaystyle\int\displaystyle\frac{xdx}{\sqrt{x^4-4hx^2+4h^2-4}}\right)+\\ \\
+{\rm C}_2\exp\left(-id_1\displaystyle\int\displaystyle\frac{xdx}{\sqrt{x^4-4hx^2+4h^2-4}}\right).
\end{array}
\end{equation*}

It is easy to see, that this function is a liouvillian function of type 2. The general solution of the equation~\eqref{63} can be represented also as follows:
\begin{equation*}
\begin{array}{l}
u\left(x\right)={\rm K}_1\sin\left(\displaystyle\frac{d_1}{2}\ln\left(x^2-2h+\sqrt{x^4-4hx^2+4h^2-4}\right)\right)+\\ \\
+{\rm K}_2\cos\left(\displaystyle\frac{d_1}{2}\ln\left(x^2-2h+\sqrt{x^4-4hx^2+4h^2-4}\right)\right),
\end{array}
\end{equation*}
where ${\rm K}_1$ and ${\rm K}_2$ are arbitrary constants.

Thus we can conclude that conditions of existence of liouvillian solutions of type 2 in the problem of motion of a heavy rigid body with a fixed point in a Hess case are formulated in Theorem~\ref{Theorem6}. When each of these conditions is fulfilled, liouvillian solutions of type 2 exist regardless of what values the other parameters of the problem take.

Let us summarize the results. In the problem of motion of a heavy rigid body with a fixed point in the Hess case there are no liouvillian solutions of type 1 and type 3. Liouvillian solutions of type 2 exist when $d_1=0$ or when $k_1=0$.

These results have been obtained in general case, when we suppose that all the nine finite poles of the function $R\left(x\right)$ defined by~\eqref{56}, are distinct. Now we will consider the special cases, when these poles can coincide with each other, i.e. the function $R\left(x\right)$ can have multiply roots.

\subsection{Description of the special cases.}

The function $R\left(x\right)$ defined by~\eqref{56} has a denominator which is the square of the product of two polynomials: the sixth degree polynomial~\eqref{57} and the cubic polynomial~\eqref{58}. Let us study now the polynomial~\eqref{57}. As we already noted, this polynomial includes only even degrees of an independent variable, as a result of which by changing $x^2=z$ we can reduce it to a third-degree polynomial with respect to $z$:
\begin{equation}\label{64}
P_{3z}=z^3-4hz^2+4\left(h^2-1\right)z+4k_1^2=0.
\end{equation}

Let us study the type of roots of the polynomial~\eqref{64}. If we put
\begin{equation*}
z=y+\frac{4h}{3}
\end{equation*}
then the polynomial~\eqref{64} takes the form
\begin{equation}\label{65}
y^3-4\left(1+\frac{h^2}{3}\right)y+4\left(k_1^2+\frac{4h^3}{27}-\frac{4h}{3}\right)=0.
\end{equation}

Thus the third-degree polynomial~\eqref{65} has the form
\begin{equation*}
y^3+3py+2q=0,
\end{equation*}
where we denote
\begin{equation*}
p=-\frac{4}{3}\left(1+\frac{h^2}{3}\right),\qquad q=2\left(k_1^2+\frac{4h^3}{27}-\frac{4h}{3}\right).
\end{equation*}

The character of the roots of the polynomial~\eqref{65} is determined by the sign of the expression
\begin{equation*}
D=q^2+p^3.
\end{equation*}

If $D>0$ then the polynomial~\eqref{65} has one real root and two complex-conjugate roots. If $D<0$ then the polynomial~\eqref{65} has three distinct real roots. If $D=0$ then the polynomial~\eqref{65} has a multiple root and all of its roots are real. In the explicit form expression $D$ can be written as follows:
\begin{equation}\label{66}
D=4\left(k_1^4+\frac{8}{27}\left(h^2-9\right)k_1^2 h-\frac{16}{27}\left(h^2-1\right)^2\right).
\end{equation}

Thus, if~\eqref{66} is zero, the polynomial~\eqref{57} has multiple roots.

We consider now the cubic polynomial~\eqref{58}. The coefficient $d_1$ is such that $d_1\in\left(-1,\, 0\right]$. Let us denote
\begin{equation*}
k_1=c^3d_1,
\end{equation*}
where $c$ is a new parameter. Then we can write the polynomial~\eqref{58} as follows:
\begin{equation}\label{67}
d_1\left(x^3-c^3\right)=0.
\end{equation}

The roots of~\eqref{67} have the following form:
\begin{equation}\label{68}
x_1=c, \quad x_2=\left(-\frac{1}{2}+\frac{i\sqrt{3}}{2}\right)c, \quad x_3=\left(-\frac{1}{2}-\frac{i\sqrt{3}}{2}\right)c.
\end{equation}

It is easy to see, that all these roots $x_1$, $x_2$, $x_3$ are distinct for $c\ne 0$. Therefore, since we do not consider the case $k_1=0$ here, the polynomial~\eqref{67} cannot have multiple roots. However the situation is possible when one of the roots~\eqref{68} of the polynomial~\eqref{67} is the root of the polynomial~\eqref{57}. Below we consider in details all possible special cases described in this Section.

\subsection{The special case of multiple roots of the polynomial~\eqref{57}.}

Let us consider the first special case when the polynomial~\eqref{57} has a multiple root. First of all we consider the case, when $d_1=0$, i.e. we consider the case of motion of the Lagrange top. For $d_1=0$ the cubic polynomial~\eqref{58} becomes a constant. Therefore in this case polynomials~\eqref{57} and~\eqref{58} have not the common roots. Further in this Section we assume, that $k_1\ne 0$.

The polynomial~\eqref{57} has multiple roots if expression $D$ defined by~\eqref{66}, is zero. This condition can be represented in the form of a biquadratic equation with respect to $k_1$:
\begin{equation}\label{69}
k_1^4+\frac{8}{27}\left(h^2-9\right)k_1^2 h-\frac{16}{27}\left(h^2-1\right)^2=0.
\end{equation}

From the equation~\eqref{69} we find the value of $k_1^2$:
\begin{equation*}
k_1^2=\frac{4}{27}\left(\left(h^2+3\right)\sqrt{h^2+3}-h\left(h^2-9\right)\right).
\end{equation*}

In this case the cubic polynomial~\eqref{64} has three real roots:
\begin{equation*}
z_1=\frac{4h}{3}-\frac{4}{3}\sqrt{h^2+3},
\end{equation*}
\begin{equation*}
z_2=z_3=\frac{4h}{3}+\frac{2}{3}\sqrt{h^2+3}.
\end{equation*}

It is easy to see that the roots $z_2$ and $z_3$ are positive. This means, that the polynomial~\eqref{57} has one real positive root of multiplicity 2 and one negative root of multiplicity 2. In addition, the polynomial~\eqref{57} has two pure imaginary roots, which are square roots of the negative expression $z_1$.

Finally in the considered special case the polynomial~\eqref{57} can be represented as follows:
\begin{equation*}
\left(x^2-z_1\right)\left(x^2-z_2\right)^2.
\end{equation*}

Moreover, the expressions $z_1$ and $z_2$ are connected by the relation
\begin{equation*}
z_2=\frac{z_1}{4}-\frac{4}{z_1}.
\end{equation*}

This relation allows us to essentially simplify the coefficients $a\left(x\right)$ and $b\left(x\right)$ of the differential equation~\eqref{45}. Indeed, we choose $z_1$ as a parameter, included in the coefficients $a\left(x\right)$ and $b\left(x\right)$. The parameters $h$ and $k_1$ are expressed in terms of $z_1$ as follows:
\begin{equation}\label{70}
h=\frac{3}{8}z_1-\frac{2}{z_1},\quad k_1=\left(\frac{2}{z_1}-\frac{z_1}{8}\right)\sqrt{-z_1}.
\end{equation}

Substitution of expressions~\eqref{70} together with the condition $d_1=0$ to the coefficients $a\left(x\right)$ and $b\left(x\right)$ of the differential equation~\eqref{45} gives the following expressions for these coefficients:
\begin{equation*}
a\left(x\right)=\frac{16z_1x^4+2\left(16-7z_1^2\right)x^2+z_1\left(z_1^2-16\right)}
{\left(x^2-z_1\right)\left(4z_1x^2+16-z_1^2\right)x},
\end{equation*}
\begin{equation*}
b\left(x\right)=\frac{z_1\left(z_1-4\right)^2\left(z_1+4\right)^2}{4\left(x^2-z_1\right)\left(4z_1x^2+16-z_1^2\right)^2x^2}.
\end{equation*}

By the formula~\eqref{48} let us find the function $R\left(x\right)$. It is the rational function which denominator has the form:
\begin{equation*}
V\left(x\right)=\left(x^2-z_1\right)^2\left(4z_1x^2-z_1^2+16\right)^2,
\end{equation*}

Thus it is easy to see, that the function $R\left(x\right)$ has four finite poles of the second order. Its partial fraction expansion has a very complicated form and we do not write it explicitly here. However it is possible to note the following properties of the partial fraction expansion of the function $R\left(x\right)$.
\begin{enumerate}
\item The coefficients $b_i$ of $\displaystyle\frac{b_i}{\left(x-x_i\right)^2}$, where $x_i$, $i=1, 2$ are roots of the polynomial
\begin{equation*}
x^2-z_1=0,
\end{equation*}
are all equal
\begin{equation*}
b_i=-\frac{3}{16}, \quad i=1, 2.
\end{equation*}

\item The coefficients $b_i$ of $\displaystyle\frac{b_i}{\left(x-x_i\right)^2}$, where $x_i$, $i=3, 4$ are roots of the polynomial
\begin{equation}\label{71}
4z_1x^2-z_1^2+16=0,
\end{equation}
are all equal
\begin{equation*}
b_i=-\frac{z_1^2+8}{2\left(3z_1^2+16\right)}, \quad i=3, 4.
\end{equation*}

\item The Laurent expansion of $R\left(x\right)$ at $x=\infty$ has the form:
\begin{equation*}
R\left(x\right)=\frac{2}{x^2}+O\left(\frac{1}{x^4}\right).
\end{equation*}
\end{enumerate}

Let us find out whether the differential equation~\eqref{59} in the considered critical case can have liouvillian solutions of type 1 (the existence of liouvillian solutions of type 2 for all values of parameters $h$ and $k_1$ was proved above). According to the Kovacic algorithm for searching liouvillian solutions of type 1, let us calculate the constants $\alpha^{\pm}_c$. For the finite poles $x=x_i$, $i=1, 2$ these constants are all equal
\begin{equation*}
\alpha_{x_i}^+=\frac{3}{4},\quad \alpha_{x_i}^-=\frac{1}{4}, \quad i=1, 2.
\end{equation*}

The constants $\alpha_{\infty}^{\pm}$ are equal
\begin{equation*}
\alpha_{\infty}^+=2,\quad \alpha_{\infty}^-=-1.
\end{equation*}

It is easy to see, that the coefficients $b_i$ of $\displaystyle\frac{b_i}{\left(x-x_i\right)^2}$, where $x=x_i$, $i=3, 4$ are the roots of the polynomial~\eqref{71}, depend on the parameter $z_1$. Let us estimate these coefficients. First of all, we can state, that $b_i<0$, $i=3, 4$. Let us prove now that the following inequality is valid for these coefficients:
\begin{equation*}
b_i<-\frac{3}{16},\quad i=3, 4.
\end{equation*}

Indeed, the inequality
\begin{equation*}
-\frac{z_1^2+8}{2\left(3z_1^2+16\right)}<-\frac{3}{16}
\end{equation*}
can be rewritten in the form:
\begin{equation*}
\frac{z_1^2+8}{3z_1^2+16}-\frac{3}{8}>0.
\end{equation*}

The latter inequality is equivalent to the inequality 
\begin{equation*}
\frac{16-z_1^2}{8\left(3z_1^2+16\right)}>0,
\end{equation*}
which is valid for all $z_1$ such that $k_1\ne 0$.

Moreover, analysis of the expressions $b_i$, $i=3, 4$ shows, that these expressions take a minimum for $z_1=0$. This minimum 
equals
\begin{equation*}
b_i=-\frac{1}{4}.
\end{equation*}

Thus, for the coefficients $b_i$, $i=3, 4$ we have the following estimate:
\begin{equation*}
-\frac{1}{4}<b_i<-\frac{3}{16}.
\end{equation*}

Then the constant
\begin{equation*}
\alpha_{x_i}^+=\frac{1}{2}+\frac{1}{2}\sqrt{1+4b_i}=\frac{1}{2}-\frac{z_1}{2\sqrt{3z_1^2+16}}, \quad i=3, 4
\end{equation*}
satisfies the estimate
\begin{equation*}
\frac{1}{2}<\alpha_{x_i}^+<\frac{3}{4},
\end{equation*}
and the constant
\begin{equation*}
\alpha_{x_i}^-=\frac{1}{2}-\frac{1}{2}\sqrt{1+4b_i}=\frac{1}{2}+\frac{z_1}{2\sqrt{3z_1^2+16}}, \quad i=3, 4
\end{equation*}
satisfies the estimate
\begin{equation*}
\frac{1}{4}<\alpha_{x_i}^-<\frac{1}{2},
\end{equation*}

Therefore, in the considered case the constant $d$, calculated by the formula~\eqref{49}, can be only zero. The value $d=0$ corresponds to the following sets of signs plus (+) and minus (-):$\,$ $s=\left(s\left(\infty\right),\,s\left(x_1\right),\,s\left(x_2\right),\,s\left(x_3\right),\,s\left(x_4\right)\right)$
\begin{equation*}
\begin{array}{l}
s_1=\left(+,\,+,\,-,\,+,\,-\right), \quad s_2=\left(+,\,-,\,+,\,+,\,-\right), \\ \\
s_3=\left(+,\,+,\,-,\,-,\,+\right),\quad s_4=\left(+,\,-,\,+,\,-,\,+\right).
\end{array}
\end{equation*}

We need to check all these sets. Let us start with the set $s_1$. For the set $s_1$ we calculate the function $\theta$ by the formula~\eqref{50}. For the set $s_1$ this function has the form:
\begin{equation*}
\theta=\frac{3}{4\left(x-x_1\right)}+\frac{1}{4\left(x-x_2\right)}+\frac{\alpha_{x_3}^+}{\left(x-x_3\right)}+
\frac{\alpha_{x_3}^-}{\left(x-x_4\right)}.
\end{equation*}

The polynomial $P$ of degree $d=0$ has the form $P\equiv 1$. The substitution of this polynomial to the differential 
equation~\eqref{51} reduces it to the equation~\eqref{62}. The substitution of the function $\theta$ to the differential equation~\eqref{62} gives in the left hand side of this equation the rational expression which vanishes only for $z_1=0$. But we assume that $z_1\ne 0$. Therefore we can state that the rational expression in the left hand side of~\eqref{62} is not zero. Thus, in the considered critical case for the set $s_1$ there are no liouvillian solutions of type 1 in the problem of motion of a heavy rigid body with a fixed point in the Hess case. The fact of nonexistence of liouvillian solutions of type 1 for the sets $s_2$, $s_3$, $s_4$ in the considered critical case is established similarly. Finally, we can state, that there are no liouvillian solutions of type 1 in the problem of motion of a heavy rigid body with a fixed point in the Hess case.

Now let us consider the case $d_1\ne 0$. Let us choose $d_1$ and $z_1$ as a parameters, included in the coefficients $a\left(x\right)$ and $b\left(x\right)$ of the differential equation~\eqref{45}. Substitution of the expressions~\eqref{70} to the coefficients $a\left(x\right)$ and $b\left(x\right)$ of the differential equation~\eqref{45} gives the following expressions for these coefficients:
\begin{equation*}
a\left(x\right)=\frac{U_1\left(x\right)}{V_1\left(x\right)},
\end{equation*}
\begin{equation*}
\begin{array}{l}
U_1\left(x\right)=32d_1z_1^2x^7+\left(z_1^2-16\right)\bigl[8d_1z_1x^5+16z_1\sqrt{-z_1}x^4-16d_1z_1^2x^3-2\left(7z_1^2-16\right)
\sqrt{-z_1}x^2+\\ \\ +z_1\left(z_1^2-16\right)\sqrt{-z_1}\bigr],
\end{array}
\end{equation*}
\begin{equation*}
V_1\left(x\right)=x\left(x^2-z_1\right)\left(4z_1x^2-z_1^2+16\right)\left(8d_1z_1x^3+\left(z_1^2-16\right)\sqrt{-z_1}\right),
\end{equation*}
\begin{equation*}
b\left(x\right)=\frac{\left(8d_1z_1x^3+16\sqrt{-z_1}-z_1^2\sqrt{-z_1}\right)\left(8d_1z_1x^3-16\sqrt{-z_1}+
z_1^2\sqrt{-z_1}\right)}{4x^2\left(x^2-z_1\right)\left(4z_1x^2+16-z_1^2\right)^2}.
\end{equation*}

By the formula~\eqref{48} let us find the function $R\left(x\right)$. It is the rational function which denominator has the form:
\begin{equation*}
V\left(x\right)=\left(x^2-z_1\right)^2\left(4z_1x^2-z_1^2+16\right)^2\left(8d_1z_1x^3-16\sqrt{-z_1}+z_1^2\sqrt{-z_1}\right)^2.
\end{equation*}

Thus, it is easy to see that the function $R\left(x\right)$ has seven finite poles of the second order. Its partial fraction expansion has a very complicated form and we do not write it explicitly here. However, it is possible to note the following properties of the partial fraction expansion of the function $R\left(x\right)$.
\begin{enumerate}
\item The coefficients $b_i$ of $\displaystyle\frac{b_i}{\left(x-x_i\right)^2}$, where $x_i$, $i=1, 2, 3$ are roots of the cubic polynomial
\begin{equation*}
8d_1z_1x^3-16\sqrt{-z_1}+z_1^2\sqrt{-z_1}=0,
\end{equation*}
are all equal
\begin{equation*}
b_i=\frac{3}{4}, \quad i=1, 2, 3.
\end{equation*}

\item The coefficients $b_i$ of $\displaystyle\frac{b_i}{\left(x-x_i\right)^2}$, where $x_i$, $i=4, 5$ are roots of the polynomial
\begin{equation*}
x^2-z_1=0,
\end{equation*}
are all equal
\begin{equation*}
b_i=-\frac{3}{16}, \quad i=4, 5.
\end{equation*}

\item The coefficients $b_i$ of $\displaystyle\frac{b_i}{\left(x-x_i\right)^2}$, where $x_i$, $i=6, 7$ are roots of the polynomial~\eqref{71} are all equal
\begin{equation*}
b_i=\frac{1}{4}\left(\frac{z_1^2\left(d_1^2-2\right)-16\left(d_1^2+1\right)}{16+3z_1^2}\right), \quad i=6, 7.
\end{equation*}

\item The Laurent expansion of $R\left(x\right)$ at $x=\infty$ has the form:
\begin{equation*}
R\left(x\right)=-\frac{\left(1+4d_1^2\right)}{4x^2}+O\left(\frac{1}{x^4}\right).
\end{equation*}
\end{enumerate}

The Laurent expansion of $R\left(x\right)$ at $x=\infty$ allows us to state that in this special case the differential equation~\eqref{45} can have liouvillian solutions of type 2 only. Moreover, it is easy to see that the coefficients $b_i$ of $x=x_i$, $i=6, 7$ depend on the parameters  $d_1$ and $z_1$. Let us estimate these coefficients. According to the restrictions on the parameter $d_1$ we can state that $b_i<0$, $i=6, 7$. Let us prove now that the following inequality is valid for these coefficients:
\begin{equation*}
b_i<-\frac{3}{16}, \quad i=6, 7.
\end{equation*}

We rewrite the latter inequality in the form:
\begin{equation}\label{72}
\frac{16\left(d_1^2+1\right)+\left(2-d_1^2\right)z_1^2}{16+3z_1^2}>\frac{3}{4}.
\end{equation}

The denominator of expression in the left hand side of~\eqref{72} is positive. Therefore we can rewrite~\eqref{72} as follows:
\begin{equation*}
16d_1^2+16+2z_1^2-d_1^2z_1^2>12+\frac{9}{4}z_1^2.
\end{equation*}

This inequality is equivalent to the inequality
\begin{equation*}
16\left(d_1^2+\frac{1}{4}\right)>z_1^2\left(d_1^2+\frac{1}{4}\right).
\end{equation*}
or
\begin{equation*}
16>z_1^2.
\end{equation*}

The latter inequality is valid for all $z_1$ such that $k_1\ne 0$.

To find liouvillian solutions of the differential equation~\eqref{45} in the considered special case we will use the Kovacic algorithm. Let us define the sets $E_{x_i}$, $i=1, \ldots, 7$ corresponding to finite poles and the set $E_{\infty}$ corresponding to the pole at $x=\infty$. For the finite poles $x=x_i$, $i=1, 2, 3$ the corresponding sets have the form:
\begin{equation*}
E_{x_i}=\{-2, 2, 6\},\quad i=1, 2, 3.
\end{equation*}

For the finite poles $x_i$, $i=4, 5$ the corresponding sets $E_{x_i}$ have the form:
\begin{equation*}
E_{x_i}=\{1, 2, 3\},\quad i=4, 5.
\end{equation*}

Taking into account estimations made for the coefficients $b_i$, $i=6, 7$, we can state that the sets $E_{x_i}$, corresponding to these poles, contain only one element and have the form
\begin{equation*}
E_{x_i}=\{2\},\quad i=6, 7.
\end{equation*}

The set $E_{\infty}$ corresponding to the pole of $R\left(x\right)$ at $x=\infty$ also contains only one element and has the form
\begin{equation*}
E_{\infty}=\{2\}.
\end{equation*}

Now we should determine the families $s=\left(e_{\infty}, e_{x_1}, e_{x_2}, e_{x_3}, e_{x_4}, e_{x_5}, e_{x_6}, e_{x_7}\right)$ for which the constant $d$ calculating by the formula~\eqref{52}, is a non-negative integer. It is easy to see, that $d$ is a non-negative integer for three families $s$ only. For the family
\begin{equation*}
s_1=\left(2, -2, -2, -2, 1, 1, 2, 2\right)
\end{equation*}
we have $d=1$ and for the families
\begin{equation*}
s_2=\left(2, -2, -2, -2, 3, 1, 2, 2\right),\quad s_3=\left(2, -2, -2, -2, 1, 3, 2, 2\right),
\end{equation*}
we have $d=0$. Let us check the family $s_1$. Using the elements of this family we calculate the function $\theta$ by the formula~\eqref{53}. For the family $s_1$ this function has the form:
\begin{equation*}
\theta=-\frac{24d_1z_1x^2}{8d_1z_1x^3-16\sqrt{-z_1}+z_1^2\sqrt{-z_1}}+\frac{x}{x^2-z_1}+\frac{8z_1x}{4z_1x^2-z_1^2+16}.
\end{equation*}

The polynomial $P$ of degree $d=1$  has the form $P=x+B$, where $B$ indefinite coefficient. This polynomial should identically satisfy differential equation~\eqref{54}. After substitution of the polynomial $P$ and the functions $\theta$ and $R\left(x\right)$ with this equation we obtain in the left hand side of~\eqref{54} the rational expression. The numerator of this expression is the 12th degree polynomial
\begin{equation*}
P_{12}=-8192d_1^3z_1^5\left(1+4d_1^2\right)Bx^{12}+\cdots
\end{equation*}

For the leading coefficient of this polynomial to vanish we put $B=0$. Then the numerator of the rational expression in the left hand side of~\eqref{54} takes the form of the 10th degree polynomial
\begin{equation*}
P_{10}=12288z_1^4d_1^4\sqrt{-z_1}\left(z_1-4\right)\left(z_1+4\right)x^{10}+\cdots
\end{equation*}

Since $z_1<0$, then $z_1-4<0$. We do not consider the cases $z_1=0$ (i.e. $k_1=0$) and $d_1=0$, because these cases have been already investigated. The in case $z_1=-4$ we have $z_2=0$, i.e. $k_1=0$. Therefore the polynomial $P_{10}$ is not identically zero. This means that for the for the family $s_1$ the differential equation~\eqref{45} does not have liouvillian solutions of type 2. Similarly we can prove that equation~\eqref{45} does not have liouvillian solutions for the families $s_2$ and $s_3$. Finally we can state the following Theorem.
\begin{thm}\label{Theorem8}
In the special case of multiple roots of the polynomial~\eqref{57} the second-order linear differential equation~\eqref{55} (or~\eqref{45}) does not have liouvillian solutions.
\end{thm}

Thus, in the first special case we have considered, when the polynomial~\eqref{57} has multiple roots, equation~\eqref{55} does not have liouvillian solutions. We turn now to the second special case when the polynomials~\eqref{57} and~\eqref{58} have a common root.

\subsection{The special case of the common root of the polynomials~\eqref{57} and~\eqref{58}.}

Let us consider now the case, when the root $x_1=c$ the real root of the polynomial~\eqref{58} from the roots~\eqref{68} is also the root of the polynomial~\eqref{57}. The parameter $k_1$ is expressed in terms of $x_1=c$ and $d_1$ as follows:
\begin{equation*}
k_1=c^3d_1.
\end{equation*}

The condition that $x_1=c$ is a root of the polynomial~\eqref{57} has the form:
\begin{equation}\label{73}
\left(1+4d_1^2\right)c^4-4hc^2+4h^2-4=0.
\end{equation}

We express $d_1$ from the condition~\eqref{73}. As a result we obtain:
\begin{equation}\label{74}
d_1^2=\frac{4-\left(c^2-2h\right)^2}{4c^4}.
\end{equation}

We will use~\eqref{74} to simplify the coefficients $a\left(x\right)$ and $b\left(x\right)$ of the linear differential equation~\eqref{45}. We will consider $c$ and $h$ as parameters included in these coefficients. Taking into account~\eqref{74}, we can write $a\left(x\right)$ and $b\left(x\right)$ as follows:
\begin{equation*}
\begin{array}{l}
a\left(x\right)=\displaystyle\frac{x^2\left(x^3-3c^3\right)}{\left(x-c\right)\left(x+c\right)\left(x^4+\left(c^2-4h\right)x^2+
c^4-4c^2h+4h^2-4\right)}-\\ \\
-\displaystyle\frac{4\left(h^2-1\right)x^3+\left(2h-c^2+2\right)\left(2h^2-c^2-2\right)c^3}{x\left(x-c\right)\left(x^2+xc+c^2\right)
\left(x^4+\left(c^2-4h\right)x^2+c^4-4c^2h+4h^2-4\right)}+\\ \\
+\displaystyle\frac{c^2x\left(12chx+4h^2-4-c^4+4c^2h\right)}{\left(x-c\right)\left(x+c\right)\left(x^2+xc+c^2\right)\left(x^4+
\left(c^2-4h\right)x^2+c^4-4c^2h+4h^2-4\right)},
\end{array}
\end{equation*}
\begin{equation*}
b\left(x\right)=-\frac{\left(x^2-xc+c^2\right)\left(x^2+xc+c^2\right)\left(2h-c^2+2\right)\left(2h-c^2-2\right)}{4c^4x^2\left(
x^4+\left(c^2-4h\right)x^2+c^4-4c^2h+4h^2-4\right)}.
\end{equation*}

Now we find the function $R\left(x\right)$ by the formula~\eqref{48}. It is the rational function which denominator has the form:
\begin{equation*}
V\left(x\right)=\left(x-c\right)^2\left(x^2+xc+c^2\right)^2\left(x+c\right)^2\left(
x^4+\left(c^2-4h\right)x^2+c^4-4c^2h+4h^2-4\right)^2c^4.
\end{equation*}

Thus, the function $R\left(x\right)$ has eight finite poles of the second order. It has the poles at $x=c$ and $x=-c$, at $x=x_i$, $i=1,\ldots, 4$, where $x_i$, $i=1,\ldots, 4$ are roots of the fourth degree polynomial
\begin{equation}\label{75}
x^4+\left(c^2-4h\right)x^2+c^4-4c^2h+4h^2-4=0,
\end{equation}
and at $x=x_i$, $i=5, 6$, where $x_i$, $i=5, 6$ are roots of the polynomial
\begin{equation}\label{76}
x^2+xc+c^2=0.
\end{equation}

The partial fraction expansion of $R\left(x\right)$ has a very complicated form and we do not write it explicitly. However it is possible to note the following properties of the partial fraction expansion of the function $R\left(x\right)$.
\begin{enumerate}
\item The coefficient $b_c$ of $\displaystyle\frac{b_c}{\left(x-c\right)^2}$ equals
\begin{equation*}
b_c=\frac{5}{16}.
\end{equation*}

\item The coefficient $b_{-c}$ of $\displaystyle\frac{b_{-c}}{\left(x+c\right)^2}$ equals
\begin{equation*}
b_{-c}=-\frac{3}{16}.
\end{equation*}

\item The coefficients $b_i$ of $\displaystyle\frac{b_i}{\left(x-x_i\right)^2}$, where $x_i$, $i=1, 2, 3, 4$ are roots of the polynomial~\eqref{75} are all equal
\begin{equation*}
b_i=-\frac{3}{16}, \quad i=1, 2, 3, 4.
\end{equation*}

\item The coefficients $b_i$ of $\displaystyle\frac{b_i}{\left(x-x_i\right)^2}$, where $x_i$, $i=5, 6$ are roots of the polynomial~\eqref{76} are all equal
\begin{equation*}
b_i=\frac{3}{4}, \quad i=5, 6.
\end{equation*}

\item The Laurent expansion of $R\left(x\right)$ at $x=\infty$ has the form:
\begin{equation*}
R\left(x\right)=-\frac{\left(1+4d_1^2\right)}{4x^2}+O\left(\frac{1}{x^4}\right).
\end{equation*}
\end{enumerate}

The Laurent expansion of $R\left(x\right)$ at $x=\infty$ allows us to state that the differential equation~\eqref{55} in the considered special case can have liouvillian solutions of type 2 only. To find these liouvillian solutions we will use the Kovacic algorithm. To find this liouvillian solutions we will use the Kovacic algorithm. Let us define the sets $E_{x_i}$, $i=1, \ldots, 8$ corresponding to finite poles and the set $E_{\infty}$ corresponding to the pole of $R\left(x\right)$ at $x=\infty$. For the finite pole $x=c$ the corresponding set has the form:
\begin{equation*}
E_c=\{-1, 2, 5\}.
\end{equation*}

For the finite pole $x=-c$ the corresponding set has the form:
\begin{equation*}
E_{-c}=\{1, 2, 3\}.
\end{equation*}

For the finite poles $x=x_i$, $i=1, 2, 3, 4$ the corresponding sets $E_{x_i}$ have the form:
\begin{equation*}
E_{x_i}=\{1, 2, 3\},\quad i=1, 2, 3, 4.
\end{equation*}

At last, for the finite poles $x=x_i$, $i=5, 6$ the corresponding sets $E_{x_i}$ have the form:
\begin{equation*}
E_{x_i}=\{-2, 2, 6\},\quad i=5, 6.
\end{equation*}

The set $E_{\infty}$ corresponding to the pole of $R\left(x\right)$ at $x=\infty$ contains only one element and has the form:
\begin{equation*}
E_{\infty}=\{2\}.
\end{equation*}

Now we should determine the families
\begin{equation*}
s=\left(e_{\infty}, e_{c}, e_{-c}, e_{x_1}, e_{x_2}, e_{x_3}, e_{x_4}, e_{x_5}, e_{x_6}\right),
\end{equation*}
for which the constant $d$ calculated by the formula~\eqref{52}, is a non-negative integer. It is easy to see, that the minimal value of the sum of the elements of sets, corresponding to finite poles, is zero. Therefore, for the family
\begin{equation*}
s_1=\left(2, -1, 1, 1, 1, 1, 1, -2, -2\right)
\end{equation*}
we have $d=1$, and for the families
\begin{equation*}
\begin{array}{l}
s_2=\left(2, -1, 3, 1, 1, 1, 1, -2, -2\right),\; s_3=\left(2, -1, 1, 3, 1, 1, 1, -2, -2\right),\\ \\
s_4=\left(2, -1, 1, 1, 3, 1, 1, -2, -2\right),\; s_5=\left(2, -1, 1, 1, 1, 3, 1, -2, -2\right),\\ \\
s_6=\left(2, -1, 1, 1, 1, 1, 3, -2, -2\right),\; s_7=\left(2, -1, 2, 2, 1, 1, 1, -2, -2\right),\\ \\
s_8=\left(2, -1, 2, 1, 2, 1, 1, -2, -2\right),\; s_9=\left(2, -1, 2, 1, 1, 2, 1, -2, -2\right),\\ \\
s_{10}=\left(2, -1, 2, 1, 1, 1, 2, -2, -2\right),\; s_{11}=\left(2, -1, 1, 2, 2, 1, 1, -2, -2\right),\\ \\
s_{12}=\left(2, -1, 1, 2, 1, 2, 1, -2, -2\right),\; s_{13}=\left(2, -1, 1, 2, 1, 1, 2, -2, -2\right),\\ \\
s_{14}=\left(2, -1, 1, 1, 2, 2, 1, -2, -2\right),\; s_{15}=\left(2, -1, 1, 1, 2, 1, 2, -2, -2\right),\\ \\
s_{16}=\left(2, -1, 1, 1, 1, 2, 2, -2, -2\right)
\end{array}
\end{equation*}
we have $d=0$. We must check all these families. We start with the family $s_1$. According to the algorithm, let us find the function $\theta$ by the formula~\eqref{53}. For the family $s_1$ this function has the form:
\begin{equation*}
\theta=\frac{2x^3+\left(c^2-4h\right)x}{x^4+\left(c^2-4h\right)x^2+c^4-4c^2h+4h^2-4}-\frac{2x+c}{x^2+xc+c^2}+
\frac{1}{2\left(x+c\right)}-\frac{1}{2\left(x-c\right)}.
\end{equation*}

The polynomial $P$ of degree $d=1$ equals $P\equiv x+B$, where $B$ is indefinite coefficient. This polynomial should identically satisfy the differential equation~\eqref{54}. After substitution of the polynomial $P=x+b$ and the functions $\theta$ and $R\left(x\right)$ to this equation, we obtain in the left hand side of~\eqref{54} the rational expression. Its numerator has the form of the 8th degree polynomial
\begin{equation*}
P_8=4\left(h^2-c^2h-1\right)Bx^{8}+\cdots
\end{equation*}

If we put $B=0$ then the polynomial $P_8$ takes the form:
\begin{equation*}
P_6=3c^3\left(2h-c^2+2\right)\left(2h-c^2-2\right)x^6+\cdots
\end{equation*}

The coefficient at $x^6$ of the polynomial $P_6$ vanishes only if $c= 0$ or $d_1=0$. None of these possibilities is considered by us; therefore, the polynomial $P_6$ does not identically vanish. We can verify similarly that the sixth degree polynomial obtained from $P_8$ does not identically vanish, if we assume that
\begin{equation}\label{77}
h^2-c^2h-1=0.
\end{equation}

Finally this means that equation~\eqref{55} has not liouvillian solutions of type 2 for the family $s_1$ in the considered special case.

Now let us consider the families $s_2, \ldots, s_6$. For the family $s_2$ the function $\theta$ calculated by the formula~\eqref{53}, has the form
\begin{equation*}
\theta=\frac{2x^3+\left(c^2-4h\right)x}{x^4+\left(c^2-4h\right)x^2+c^4-4c^2h+4h^2-4}-\frac{2x+c}{x^2+xc+c^2}+
\frac{3}{2\left(x+c\right)}-\frac{1}{2\left(x-c\right)}.
\end{equation*}

The polynomial $P$ of degree $d=0$ has the form $P\equiv 1$. This polynomial should identically satisfy differential equation~\eqref{54}. After substitution of $P=1$ and the functions $\theta$ and $R\left(x\right)$ to the differential equation~\eqref{54} we obtain in the left hand side of this equation the rational expression. Its numerator has the form of 8th degree polynomial
\begin{equation*}
P_8=4\left(h^2-c^2h-1\right)x^{8}+\cdots
\end{equation*}

If we assume that condition~\eqref{77} is valid, then the substitution of this condition with $P_8$ reduces it to the polynomial of the 6th degree, which is not zero for any $c$ and $d_1$ such that $c\ne 0$ and $d_1\ne 0$. Thus, the differential equation~\eqref{55} has not liouvillian solutions of type 2 for the family $s_2$. Similarly we can check the families $s_3, \ldots, s_{16}$. For all these families the differential equation~\eqref{55} has not liouvillian solutions of type 2. Thus we can state that in the special case when the polynomials~\eqref{57} and~\eqref{58} have a common root, which is not the a multiple root of the polynomial~\eqref{57}, the differential equation~\eqref{55} has not liouvillian solutions.

\subsection{The special case of the multiple common root of the polynomials~\eqref{57} and~\eqref{58}.}

Now let us consider the case, when the root $x_1=c$ -- the real root of the polynomial~\eqref{58} from the roots~\eqref{68} id the multiple root of the polynomial~\eqref{57}. In this case the cubic polynomial~\eqref{64} obtained from the polynomial~\eqref{57}, has the form
\begin{equation*}
P_{3z}=z^3-4hz^2+4\left(h^2-1\right)z+4d_1^2c^6=\left(z-c^2\right)^2\left(z+z_1\right).
\end{equation*}

Comparing the coefficients at the same powers of $z$ in two different representations of the polynomial $P_{3z}$ we obtain the following system of equations for the parameters of the problem:
\begin{equation}\label{78}
\begin{array}{l}
2c^2-z_1-4h=0,\\ \\
2c^2z_1-c^4+4h^2-4=0,\\ \\
4d_1^2c^6-c^4z_1=0.
\end{array}
\end{equation}

Solving the system~\eqref{78} we find the following expressions for the parameters $h$, $z_1$ e $d_1$ in terms of the parameter $c$:
\begin{equation}\label{79}
\begin{array}{l}
d_1^2=\displaystyle\frac{\sqrt{c^4+4}}{2c^2}-\displaystyle\frac{1}{2}, \\ \\
z_1=4d_1^2c^2=2\sqrt{c^4+4}-2c^2, \\ \\
h=c^2\left(\displaystyle\frac{1}{2}-d_1^2\right)=c^2\left(1-\displaystyle\frac{\sqrt{c^4+4}}{2c^2}\right)=c^2-
\displaystyle\frac{\sqrt{c^4+4}}{2}.
\end{array}
\end{equation}

We will use~\eqref{79} to simplify the coefficients $a\left(x\right)$ and $b\left(x\right)$ of the differential equation~\eqref{45}. Taking into account~\eqref{79} we can write the coefficients $a\left(x\right)$ and $b\left(x\right)$ of the second order linear differential equation~\eqref{45} as follows:
\begin{equation*}
a\left(x\right)=\frac{x^5+2cx^4+4c^2x^3+2c^3x^2+2c^2\left(2x+c\right)\left(\sqrt{c^4+4}-c^2\right)}{x\left(x+c\right)
\left(x^2+xc+c^2\right)\left(x^2+2\sqrt{c^4+4}-2c^2\right)},
\end{equation*}
\begin{equation*}
b\left(x\right)=\frac{\left(x^2+xc+c^2\right)\left(x^2-xc+c^2\right)\left(\sqrt{c^4+4}-c^2\right)}{2c^2x^2\left(x-c\right)
\left(x+c\right)\left(x^2+2\sqrt{c^4+4}-2c^2\right)}.
\end{equation*}

Now we find the function $R\left(x\right)$ for this special case by the formula~\eqref{48}. It is the rational function of the rather complicated form which denominator can be written as follows:
\begin{equation*}
V\left(x\right)=4c^2\left(x+c\right)^2\left(x^2+xc+c^2\right)^2\left(x^2+2\sqrt{c^4+4}-2c^2\right)^2\left(x-c\right).
\end{equation*}

Thus, the function $R\left(x\right)$ has six finite poles. At $x=c$ function $R\left(x\right)$ has the first order pole. At $x=-c$ it has the second order pole. The function $R\left(x\right)$ has also the second order poles at $x=x_i$, $i=1, 2, 3, 4$, where $x_1$ and $x_2$ are roots of the polynomial
\begin{equation}\label{80}
x^2+xc+c^2=0,
\end{equation}
and $x_3$ and $x_4$ are roots of the polynomial
\begin{equation}\label{81}
x^2+2\sqrt{c^4+4}-2c^2=0.
\end{equation}

The partial fraction expansion of $R\left(x\right)$ has a very complicated form and we do not write it here explicitly. However it is possible to note the following properties of the partial fraction expansion of the function $R\left(x\right)$.
\begin{enumerate}
\item The coefficients $b_i$ of $\displaystyle\frac{b_i}{\left(x-x_i\right)^2}$, where $x_i$, $i=1, 2$ are roots of the polynomial~\eqref{80} equal
\begin{equation*}
b_i=\frac{3}{4}, \quad i=1, 2.
\end{equation*}

\item The coefficients $b_i$ of $\displaystyle\frac{b_i}{\left(x-x_i\right)^2}$, where $x_i$, $i=3, 4$ are roots of the polynomial~\eqref{81} equal
\begin{equation*}
b_i=-\frac{3}{16}, \quad i=3, 4.
\end{equation*}

\item The coefficient $b_{-c}$ of $\displaystyle\frac{b_{-c}}{\left(x+c\right)^2}$ equals
\begin{equation*}
b_{-c}=-\frac{1}{4}.
\end{equation*}

\item The Laurent expansion of $R\left(x\right)$ at $x=\infty$ has the form:
\begin{equation*}
R\left(x\right)=-\frac{\left(1+4d_1^2\right)}{4x^2}+O\left(\frac{1}{x^4}\right),
\end{equation*}
where parameter $d_1$ are expressed in terms of $c$ by~\eqref{79}.
\end{enumerate}

The Laurent expansion of $R\left(x\right)$ at $x=\infty$ allows us to state that the differential equation~\eqref{55} in the considered special case can have liouvillian solutions of type 2 only. To find these liouvillian solutions, we will use the Kovacic algorithm. According to the algorithm, let us find the sets $E_{x_i}$, $i=1, 2,\ldots, 6$, corresponding to the finite poles and the set $E_{\infty}$ corresponding to the pole of $R\left(x\right)$ at $x=\infty$. For the first order finite pole $x=c$ the corresponding set has the form:
\begin{equation*}
E_c=\{4\}.
\end{equation*}

For the finite pole of the second order $x=-c$ the corresponding set has the form:
\begin{equation*}
E_{-c}=\{2\}.
\end{equation*}

For the finite poles of the second order $x=x_i$, $i=1, 2$ the corresponding sets $E_{x_i}$ have the form:
\begin{equation*}
E_{x_i}=\{-2, 2, 6\},\quad i=1, 2.
\end{equation*}

At last, for the finite poles of the second order $x=x_i$, $i=3, 4$ the corresponding sets $E_{x_i}$ have the form
\begin{equation*}
E_{x_i}=\{1, 2, 3\},\quad i=3, 4.
\end{equation*}

The set $E_{\infty}$ corresponding to the pole of $R\left(x\right)$ at $x=\infty$ contains only one element and has the form:
\begin{equation*}
E_{\infty}=\{2\}.
\end{equation*}

Now we should determine the families
\begin{equation*}
s=\left(e_{\infty}, e_{c}, e_{-c}, e_{x_1}, e_{x_2}, e_{x_3}, e_{x_4}, e_{x_5}, e_{x_6}\right),
\end{equation*}
for which the constant $d$ calculated by the formula~\eqref{52}, is a non-negative integer. It is easy to see, that the minimal value of the sum of elements of sets, corresponding to finite poles equals 4. This means that there are no families $s$, for which the constant $d$, calculated by~\eqref{52}, is a non-negative integer. Thus, in the considered special case the differential equation~\eqref{55} has no liouvillian solutions.

\subsection{The special case of the common non-real root of the polynomials~\eqref{57} and~\eqref{58}.}

Now let us consider the case when one of the roots $x_2$ or $x_3$ from the roots~\eqref{68} of the cubic polynomial~\eqref{58} is the root of the polynomial~\eqref{57}. If we substitute any of these roots to~\eqref{57} and take into account that $k_1=c^3d_1$, we can find, that parameters of the problem should satisfy the system of two conditions:
\begin{equation}\label{82}
h^2+hc^2-1=0, \quad \left(1+4d_1^2\right)c^4+2hc^2-2\left(h^2-1\right)=0.
\end{equation}

Let us consider the first of these conditions~\eqref{82}. This is quadratic equation with respect to $h$. Its solutions have the following form:
\begin{equation*}
h_1=\frac{1}{2}\sqrt{c^4+4}-\frac{c^2}{2}, \quad h_2=-\frac{1}{2}\sqrt{c^4+4}-\frac{c^2}{2}.
\end{equation*}

It is easy to see that for $h_2$ the following inequality holds:
\begin{equation*}
-\frac{1}{2}\sqrt{c^4+4}-\frac{c^2}{2}\leq -1,
\end{equation*}

Taking into account the range of the parameter $h$ we obtain that $h$ can take only a value $h=h_1$. Thus, we have
\begin{equation*}
h=\frac{\sqrt{c^4+4}}{2}-\frac{c^2}{2}.
\end{equation*}

Substituting this expression for $h$ to the second equation of the system~\eqref{82} and solving it with respect to $d_1^2$, we obtain
\begin{equation}\label{83}
d_1^2=-\frac{2\sqrt{c^4+4}-c^2}{4c^2}.
\end{equation}

The right hand side of the equation~\eqref{83} is negative for all values of $c\ne 0$. This means that the condition~\eqref{83} cannot be valid for all values of parameters $c$ and $d_1$. Thus, the polynomials~\eqref{57} and~\eqref{58} cannot have the common non-real root. This means, that the considered special case does not hold.

\subsection{Conclusions}
In this paper we considered the problem of motion of a heavy rigid body with a fixed point in the Hess case. The integration of this problem is reduced to solving the second order linear differential equation. We reduce this equation to the equation with rational coefficients~\eqref{45}. Using the Kovacic algorithm we studied the problem of existence of luouvillian solutions of the differential equation~\eqref{45}. We proved that this equation has liouvillian solutions of type 2 only and only if the moving rigid body is the Lagrange top $\left(d_1=0\right)$ or if the constant of the area integral is zero $\left(k_1=0\right)$. Thus the problem of existence of liouvillian solutions for the second order linear differential equation~\eqref{45}, the integration of which solves the problem of motion of a heavy rigid body with a fixed point in the Hess case, is completely solved.

\subsection{Appendix. Motion of a heavy rigid body with a fixed point in a Hess case, for which the conditions~\eqref{44} holds.}

Let us find out the motion of the Hess top in the case,  when  the conditions ~\eqref {44} holds in addition to the conditions~\eqref{7}. Condition $d_1=0$ is equivalent to the condition
\begin{equation}\label{84}
\left(A_1-A_2\right)x_1x_2=0.
\end{equation}

Further we will assume that $A_1\ne A_2$. The condition $A_1=A_2$ together with~\eqref{7} corresponds to the case of kinetic symmetry. Then~\eqref{84} is equivalent to $x_1=0$ or $x_2=0$. We will consider the case $x_1=0$ (the case $x_2=0$ is studied similarly). The condition $x_1=0$ together with~\eqref{7}, corresponds to the Lagrange integrable case
\begin{equation}\label{85}
x_1=0,\quad x_3=0,\quad A_1=A_3,\quad x_2=a\ne 0.
\end{equation}

The Hess integral~\eqref{8} takes the form
\begin{equation}\label{86}
\omega_2=0,
\end{equation}
and the area integral can be rewritten as follows:
\begin{equation}\label{87}
\omega_1\gamma_1+\omega_3\gamma_3=0,
\end{equation}
where we taking into account that the constant of this integral is zero.

The Euler equations~\eqref{5} under conditions~\eqref{85} takes the form:
\begin{equation*}
A_1\dot{\omega}_1=-Mga\gamma_3, \quad A_1\dot{\omega}_3=Mga\gamma_1.
\end{equation*}

If we introduce the standard Euler angles by the formulae
\begin{equation*}
\gamma_1=\sin\theta\cos\varphi,\quad \gamma_2=\cos\theta,\quad \gamma_3=\sin\theta\sin\varphi,
\end{equation*}
\begin{equation*}
\omega_1=\dot{\psi}\sin\theta\cos\varphi-\dot{\theta}\sin\varphi,\quad \omega_2=\dot{\psi}\cos\theta+\dot{\varphi},\quad
\omega_3=\dot{\psi}\sin\theta\sin\varphi+\dot{\theta}\cos\varphi
\end{equation*}
we can rewrite equation~\eqref{86} in the form
\begin{equation}\label{88}
\dot{\psi}\cos\theta+\dot{\varphi}=0,
\end{equation}
and the equation~\eqref{87} takes the form:
\begin{equation}\label{89}
\dot{\psi}\sin^2\theta=0.
\end{equation}

We will consider the case $\sin\theta\ne 0$ (if $\theta=0$, then all the components $\omega_1$, $\omega_2$, and $\omega_3$ of angular velocity are zero). Then, from the condition~\eqref{89} we have
\begin{equation*}
\dot{\psi}=0.
\end{equation*}

Condition~\eqref{88} gives
\begin{equation*}
\dot{\varphi}=0.
\end{equation*}

This means that for the considered motions of the Lagrange top we have
\begin{equation*}
\psi=\psi_0={\rm const}, \quad \varphi=\varphi_0={\rm const}.
\end{equation*}

Thus, the components $\omega_1$ and $\omega_3$ of angular velocity of the body takes the form:
\begin{equation*}
\omega_1=-\dot{\theta}\sin\varphi_0,\quad \omega_3=\dot{\theta}\cos\varphi_0.
\end{equation*}

The Euler equations~\eqref{5} are reduced to the differential equation
\begin{equation*}
\ddot{\theta}-\frac{Mga}{A_1}\sin\theta=0,
\end{equation*}
which describes the pendulum nutational oscillations of the Lagrange top in the vicinity of $\theta=\pi$. Thus the motions of a heavy rigid body with a fixed point in the Hess case for which $d_1=0$ and $k_1=0$ correspond to the nutational oscillations of the body.

\subsection*{Acknowledgements.}
This work was supported financially by the Russian Foundation for Basic Re\-se\-ar\-ches (grants no. 19-01-00140, 20-01-00637).


\begin{thebibliography}{99}
\bibitem{Appelroth}
G.~Appel'roth. "About the first paragraph of the paper S.~Kowalevski: Sur le probl\'eme de la rotation d’un
corps solide autour d’un point fixe (Acta Mathematica. 12:2)", \textit{Mathem. Sb.} [in Russian], {\bf 16}, No.~3, 483--507 (1892).

\bibitem{Arhangelskii}
Yu.~A.~Arkhangelskii, \textit{Rigid body analytic dynamics} [in Russian], Nauka, Moscow (1977).

\bibitem{Gashenenko}
I.~N.~Gashenenko, "The Poinsot kinematic representation of the motion of a body in the Hess case", in: Rigid-Body Mechanics [in Russian], Kiev (2010), pp.~12--20.

\bibitem{Hess} %
W.~Hess, "Ueber die Euler'schen Bewegungsgleichungen und \"uber eine neue partikul\"are L\"osung des Problems der Bewegung eines starren K\"orpers um einen festen Punkt", \textit{Math. Ann.}, {\bf 37}, No.~2, 153--181 (1890).

\bibitem{Zhukowski}
N.~Joukowsky, "Geometrische Interpretation des Hess'schen Falles der Bewegung eines schweren starren K\"orpers um einen festen Punkt", \textit{Jahresbericht der Deutschen Mathematiker-Vereinigung.}, {\bf 3}, No.~9, 62--70 (1894).

\bibitem{Kharlamov1}
P.~V.~Kharlamov, "Kinematic interpretation of the motion of a body with a fixed point," \textit{J. Appl. Maths. Mechs.}, {\bf 28}, No.~3, 615--621 (1964).

\bibitem{Kharlamov2}
P.~V.~Kharlamov, \textit{Lectures on the Rigid Body Dynamics} [in Russian], Novosibirsk University, Novosibirsk (1965).

\bibitem{Kovacic} %
J.~Kovacic, "An algorithm for solving second order linear homogeneous differential equations," \textit{J. Symb. Comput.}, {\bf 2}, 3--43 (1986).

\bibitem{Kovalev1}
A.~M.~Kovalev, "The moving angular velocity hodograph in Hess' solution of the problem of motion of a body with a fixed point," \textit{J. Appl. Maths. Mechs.}, {\bf 32}, No.~6, 1129--1137 (1968).

\bibitem{Kovalev2}
A.~M.~Kovalev, "On the motion of a body in the case of Hess," in: Rigid-Body Mechanics [in Russian], Kiev (1969), pp.~12--27.

\bibitem{Kovalev3}
A.~M.~Kovalev and V.~V.~Kirichenko, "Hodograph of the kinetic moment vector in the Hess solution," in: Rigid-Body Mechanics [in Russian], Kiev (2004), pp.~9--20.

\bibitem{Kowalevski1}
S.~Kowalevski, "Sur le probleme de la rotation d'un corps solide autour d'un point fixe," \textit{Acta Math.}, {\bf 12}, No.~1, 177--232 (1889).

\bibitem{Kowalevski2}
S.~Kowalevski, "Sur une propri\'et\'e du syst\'eme d'\'equations diff\'erentielles qui d\'efinit la rotation d'un corps solide autour d'un point fixe," \textit{Acta Math.}, {\bf 14}, No.~1, 81--93 (1890).

\bibitem{Kozlov}
V.~V.~Kozlov, \textit{Methods of qualitative analysis in the dynamics of a rigid body} [in Russian], Moscow State University, Moscow (1980).

\bibitem{Liouville}
R.~Liouville, "Sur la rotation des solides," \textit{C. R. Acad. Sci.}, {\bf 120}, No.~17, 903--906 (1895).

\bibitem{Mlodzeevski}
B.~K.~Mlodzeevskii, P.~A.~Nekrasov, "On conditions for the existence of asymptotic periodic motions in Hess's problem", \textit{Trudy Otdel. Fiz. Nauk Obschestva Lyubit. Estestvozn.} [in Russian], {\bf 6}, No.~1, 43--52 (1893).

\bibitem{Nekrasov1} %
P.~A.~Nekrasov, "On the problem of motion of a heavy rigid body about a fixed point", \textit{Mathem. Sb.} [in Russian], {\bf 16}, No.~2, 508--517 (1892).

\bibitem{Nekrasov2} %
P.~A.~Nekrasov, "Recherches analytiques sur un cas de rotation d'un solide pesant autour d'un point fixe", \textit{Math. Ann.}, {\bf 47}, 445--530 (1896).

\bibitem{Novikov}
M.~A.~Novikov, "On Stationary Motions of a Rigid Body under the Partial Hess Integral Existence", \textit{Mechanics of Solids}, {\bf 53}, No.~3, 262--270 (2018).

\bibitem{ZaitsevPolyanin}
A.~D.~Polyanin, V.~F.~Zaitsev, \textit{Handbook of Exact Solutions for Ordinary Differential Equations}, CRC Press, Boca Raton--New York (2003).

\bibitem{SyngeGriffith}
J.~L.~Synge, B.~A.~Griffith, \textit{Principles of Mechanics}, McGraw--Hill, New York, Toronto, London (1949).

\end{thebibliography}
\end{document}